\input harvmac
\overfullrule=0pt
\font\authorfont=cmcsc10 \ifx\answ\bigans\else scaled\magstep1\fi
{\divide\baselineskip by 4
\multiply\baselineskip by 3
\def\prenomat{\matrix{\hbox{hep-th/9603136}&\cr \qquad\hbox{SWAT/104}&\cr
\qquad\hbox{DTP/96/30}&\cr\hbox{LA-UR-96-1157}&\cr}}
\Title{$\prenomat$}{\vbox{\centerline{Multi-Instanton Calculus} 
\vskip2pt
\centerline{in $N=2$ Supersymmetric Gauge Theory}}}
\centerline{\authorfont Nicholas Dorey}
\bigskip
\centerline{\sl Physics Department, University College of Swansea}
\centerline{\sl Swansea SA2$\,$8PP UK $\quad$ \tt n.dorey@swansea.ac.uk}
\bigskip
\centerline{\authorfont Valentin V. Khoze}
\bigskip
\centerline{\sl Department of Physics, Centre for Particle Theory, 
University of Durham}
\centerline{\sl Durham DH1$\,$3LE UK $\quad$ \tt valya.khoze@durham.ac.uk}
\bigskip
\centerline{and}
\bigskip
\centerline{\authorfont Michael P. Mattis}
\bigskip
\centerline{\sl Theoretical Division T-8, Los Alamos National Laboratory}
\centerline{\sl Los Alamos, NM 87545 USA$\quad$ \tt mattis@pion.lanl.gov}
\vskip .3in
\noindent
The Seiberg-Witten solution of $N=2$ supersymmetric $SU(2)$ gauge theory
may be viewed as a prediction for the infinite family of constants
$\{{\cal F}_n\}$ measuring the $n$-instanton contribution
to the prepotential ${\cal F}$.
Here we examine the instanton physics
directly, in particular the contribution of the general 
self-dual solution of topological charge $n$ 
constructed by Atiyah, Drinfeld, Hitchin and Manin (ADHM).
In both the bosonic and supersymmetric cases,
we determine both the large- and short-distance behavior of all the
fields in this background. This allows us to construct the exact
classical interaction between $n$ ADHM (super-)instantons
mediated by the adjoint Higgs. We 
calculate the one- and two-instanton contributions to the low-energy
Seiberg-Witten effective action, and find precise agreement with 
their  predicted values of ${\cal F}_{1}$ and ${\cal F}_{2}$.
\vskip .2in
\centerline{Physical Review D (in press)}
\vskip .1in
\vfil\break
}
\def\etal{{\rm et al.}}
\lref\Gates{J. Gates, Nucl. Phys. B238 (1984) 349.}
\lref\dkmtwo{N. Dorey, V. Khoze and M.P. Mattis, work in preparation.}
\lref\Fayet{See for example P. Fayet and S. Ferrara, Phys.~Rep.~32 (1977)
249, Sec.~2.2.}
\lref\Jack{I. Jack, Nucl. Phys. B174 (1980) 526.}
\lref\GSW{R. Grimm, M. Sohnius and J. Wess, Nucl. Phys. B133 (1978) 275.}
\lref\BPST{A. Belavin, A. Polyakov, A. Schwartz and
Y. Tyupin, Phys. Lett. 59B (1975) 85.}
\lref\wessbagger{J. Wess and J. Bagger, {\it Supersymmetry and supergravity}, 
Princeton University Press, 1992.} 
\lref\Amati{A comprehensive review may be found in,
D. Amati, K. Konishi, Y. Meurice, G. Rossi and G. Veneziano,
Phys. Rep. 162 (1988) 169.}
\lref\Clark{T. E. Clark, O. Piguet and K. Sibold, Nucl. Phys. B143
(1978) 445.}
\lref\SWone{N. Seiberg and E. Witten, 
{\it Electric-magnetic duality, monopole
condensation, and confinement in $N=2$ supersymmetric Yang-Mills theory}, 
Nucl. Phys. B426 (1994) 19, (E) B430 (1994) 485  hep-th/9407087}
\lref\SWtwo{
N. Seiberg and E. Witten, 
{\it Monopoles, duality and chiral symmetry breaking
in $N=2$ supersymmetric QCD}, 
Nucl. Phys B431 (1994) 484 ,  hep-th/9408099}
\lref\KLTYone{A. Klemm, W. Lerche, S. Theisen and S. Yankielowicz 
{\it Simple singularities and $N=2$ supersymmetric Yang-Mills theory},  
Phys. Lett. B344 (1995) 169, hep-th/9411048}
\lref\AFone{P.C. Argyres and A.E. Faraggi, 
{\it The vacuum structure and spectrum of $N=2$ 
supersymmetric $SU(N)$ gauge theory},
 Phys. Rev. Lett. 74 (1995) 3931 , hep-th/9411057.}
\lref\KLTtwo{A. Klemm, W. Lerche and S. Theisen, 
{\it Nonperturbative effective actions of 
$N=2$ supersymmetric gauge theories}, 
CERN-TH/95-104, hep-th/9505150.}
\lref\DS{ M. Douglas and S. Shenker,
{\it Dynamics of $SU(N)$ supersymmetric gauge theory},
Nucl. Phys. B447 (1995) 271, hep-th/9503163; \hfil\break
U.H. Danielsson and B. Sundborg,
{\it The moduli space and monodromies of $N=2$ supersymmetric $SO(2R+1)$
Yang-Mills theory},
Phys. Lett. B358 (95) 273,  hep-th/9504102; \hfil\break 
A. Brandhuber and K. Landsteiner,
 {\it On the monodromies of $N=2$ supersymmetric $SO(2N)$}, 
Phys. Lett. B358 (1995) 73, hep-th/9507008; \hfil\break   
A. Hanany and Y. Oz,
{\it On the quantum moduli space of vacua of $N=2$ supersymmetric
$SU(N)$ gauge theories},
Nucl. Phys. B452 (1995) 283, hep-th/9505075;\hfil\break 
P. Argyres, M. Plesser and A. Shapere,
{\it The Coulomb phase of $N=2$ supersymmetric QCD},
 Phys. Rev. Lett. 75 (1995) 1699    hep-th/9505100; \hfil\break
P. Argyres and A. Shapere,
{\it The vacuum structure of N=2 super QCD with
 classical gauge groups},  RU-95-61,  hep-th/9505175; \hfil\break
A. Hanany,
{\it On the quantum moduli space of N=2
 supersymmetric gauge theories}, IASSNS-HEP-95/76,   hep-th/9505176. }
\lref\BIone{For reviews of recent progress in the field, see:\hfil\break
A. Bilal, {\it Duality in $N=2$ SUSY $SU(2)$ Yang-Mills theory:
A pedagogical introduction to the work of Seiberg and Witten}, 
LPTENS-95-53, hep-th/9601007; \hfil\break C. Gomez and R. Hernandez, 
{\it Electric-magnetic duality and effective field theories}, 
FTUAM 95/36,  hep-th/9510023.}
\lref\MAtone{
M. Matone, {\it Instantons and recursion relations in $N=2$ SUSY gauge theory}
 Phys. Lett. B357 (1995) 342,    hep-th/9506102.  }
\lref\FPone{ D. Finnell and P. Pouliot,
{\it Instanton calculations versus exact results in 4 dimensional 
SUSY gauge theories},
Nucl. Phys. B453 (95) 225, hep-th/9503115. }
\lref\ISone{ K.Ito and N. Sasakura,    
{\it One instanton calculations in $N=2$ supersymmetric $SU(N_c)$
Yang-Mills theory},
KEK-TH-470, hep-th/9602073.       }
\lref\FBone{ F. Ferrari and A. Bilal,    
{\it The strong coupling spectrum of the Seiberg-Witten theory},
LPTENS-96-16,  hep-th/9602082.}
\lref\Sone{ N. Seiberg, Phys. Lett. B206 (1988) 75. }
\lref\ADSone{ I. Affleck, M. Dine and N. Seiberg, Nucl. Phys. B241
(1984) 493; Nucl. Phys. B256 (1985) 557.  }
\lref\Aone{I. Affleck, Nucl. Phys. B191 (1981) 429.}
\lref\Cone{ S. Cordes, Nucl. Phys. B273 (1986) 629.}
\lref\NSVZone{ V. A. Novikov, M. A. Shifman, A. I. Vainshtein and
V. I. Zakharov, Nucl Phys. B229 (1983) 394; Nucl. Phys. B229 (1983)
407; Nucl. Phys. B260 (1985) 157. }
\lref\FSone{ J. Fuchs and M. G. Schmidt, Z. Phys. C30 (1986) 161. }
\lref\FUone{ J. Fuchs, Nucl Phys B272 (1986) 677; Nucl Phys B 282
(1987) 437. }
\lref\ADHMone{ M. F. Atiyah, V. G. Drinfeld, N. J. Hitchin and
Yu. I. Manin, Phys. Lett. A65 (1978) 185. }
\lref\ADHMtwo{ V. G. Drinfeld and Yu. I. Manin, Commun. Math. Phys. 
63 (1978) 177.} 
\lref\Oone{ H. Osborn, Ann. Phys. 135 (1981) 373. }
\lref\CGTone{ E. Corrigan, P. Goddard and S. Templeton,
Nucl. Phys. B151 (1979) 93; \hfil\break
   E. Corrigan, D. Fairlie, P. Goddard and S. Templeton,
    Nucl. Phys. B140 (1978) 31.}
\lref\CGone{ E. Corrigan and P. Goddard, {\it Some aspects of
instantons}, DAMTP 79/80, published in
the Proceedings of \it Geometrical and Topological Methods in Gauge 
Theories\rm,  Montreal 1979, eds. J.P. Harnad and S. Shnider,
Springer Lecture Notes in Physics 129 (Springer Verlag 1980).}
\lref\CWSone{ N. H. Christ, E. J. Weinberg and N. K. Stanton, Phys
Rev D18 (1978) 2013. }
\lref\GMOone{ P. Goddard, P. Mansfield and H. Osborn, Phys. Lett. B98 
(1981) 59. }
\lref\MAone{ P. Mansfield, Nucl. Phys. B186 (1981) 287. }
\lref\JNRone{ R. Jackiw, C. Nohl and C. Rebbi, Phys. Rev. D15 (1979)
1642.  }
\lref\Olive{C. Montonen and D. Olive, Phys. Lett. 72B (1977) 117.} 
\lref\yung{A. Yung, {\it Instanton-induced effective
Lagrangian in the Seiberg-Witten model}, hep-th/9605096.}
\lref\tHooft{G. 't Hooft, Phys. Rev. D14 (1976) 3432; ibid.
D18 (1978) 2199.}
\lref\Derrick{G. H. Derrick,  J. Math.~Phys.~5 \rm (1964) 1252;
R. Hobart,  Proc.~Royal.~Soc. London  82 \rm (1963) 201.}
\def\bose{{\rm bose}}
\def\rmD{{\rm D}}
\def\Nequalstwo{\Psi}
\def\eff{{\rm eff}}
\def\inst{{\rm inst}}
\def\higgs{{\rm higgs}}
\def\yuk{{\rm yuk}}
\def\fermi{{\rm fermi}}
\def\trtwo{\tr^{}_2\,}
\def\finv{f^{-1}}
\def\Ubar{\bar U}
\def\wbar{\bar w}
\def\fbar{\bar f}
\def\abar{\bar a}
\def\bbar{\bar b}
\def\Deltabar{\bar\Delta}
\def\dalpha{{\dot\alpha}}
\def\dbeta{{\dot\beta}}
\def\dgamma{{\dot\gamma}}
\def\ddelta{{\dot\delta}}
\def\Sbar{\bar S}
\def\Im{{\rm Im}}
\def\susytwodef{A.1}
\def\useful{A.2}
\def\moreuseful{A.3}
\def\vmntrans{A.4}
\def\xiofxdef{A.5}
\def\sconfdef{A.6}
\def\example{A.7}
\def\sst{\scriptscriptstyle}
\def\cld{C_{\sst\rm LD}^{}}
\def\csd{C_{\sst\rm SD}^{}}
\def\bigI{{\rm I}_{\sst 3\rm D}}
\def\Mr{{\rm M}_{\sst R}}
\def\cJ{C_{\sst J}}
\def\one{{\sst(1)}}
\def\two{{\sst(2)}}
\def\vsd{v^{\sst\rm SD}}
\def\vasd{v^{\sst\rm ASD}}
\def\Phibar{\bar\Phi}
\def\F{{\cal F}_{\sst\rm SW}}
\def\P{{\cal P}}
\def\A{{\cal A}}
\def\susy{supersymmetry}
\def\sigmabar{\bar\sigma}
\def\barsigma{\sigmabar}

\def\cl{{\,\rm cl}}
\def\lambdabar{\bar\lambda}
\def\R{{R}}
\def\psibar{\bar\psi}
\def\sqrtwo{\sqrt{2}\,}
\def\etabar{\bar\eta}
\def\Thetabar{{\bar\Theta_0}}
\def\Qbar{\bar Q}
\def\susic{supersymmetric}
\def\vhiggs{{\rm v}}
\def\vhiggsa{{\cal A}_{\sst00}}
\def\vbarhiggs{\bar{\rm v}}

\def\novetal{Novikov et al.}
\def\Novetal{Novikov et al.}

\def\Abar{A^\dagger}

\def\zero{{\scriptscriptstyle(0)}}
\def\new{{\scriptscriptstyle\rm new}}

\def\uA{\,\lower 1.2ex\hbox{$\sim$}\mkern-13.5mu A}
\def\uX{\,\lower 1.2ex\hbox{$\sim$}\mkern-13.5mu X}
\def\uD{\,\lower 1.2ex\hbox{$\sim$}\mkern-13.5mu {\rm D}}

\def\uAzero{{\uA}^\zero}
\def\upsizero{{\upsi}^\zero}
\def\uF{\,\lower 1.2ex\hbox{$\sim$}\mkern-13.5mu F}
\def\uW{\,\lower 1.2ex\hbox{$\sim$}\mkern-13.5mu W}
\def\uWbar{\,\lower 1.2ex\hbox{$\sim$}\mkern-13.5mu {\overline W}}

\def\uAbar{{\uA}^\dagger}
\def\uAbarzero{{\uA}^{\dagger\zero}}

\def\uFbar{{\uF}^\dagger}

\def\uV{\,\lower 1.2ex\hbox{$\sim$}\mkern-13.5mu V}
\def\uv{\,\lower 1.0ex\hbox{$\scriptstyle\sim$}\mkern-11.0mu v}
\def\uPsi{\,\lower 1.2ex\hbox{$\sim$}\mkern-13.5mu \Psi}
\def\uPhi{\,\lower 1.2ex\hbox{$\sim$}\mkern-13.5mu \Phi}
\def\uchi{\,\lower 1.5ex\hbox{$\sim$}\mkern-13.5mu \chi}
\def\Psibar{\bar\Psi}
\def\uPsibar{\,\lower 1.2ex\hbox{$\sim$}\mkern-13.5mu \Psibar}
\def\upsi{\,\lower 1.5ex\hbox{$\sim$}\mkern-13.5mu \psi}
\def\psibar{\bar\psi}
\def\upsibar{\,\lower 1.5ex\hbox{$\sim$}\mkern-13.5mu \psibar}
\def\upsibarzero{\,\lower 1.5ex\hbox{$\sim$}\mkern-13.5mu \psibar^\zero}
\def\ulambda{\,\lower 1.2ex\hbox{$\sim$}\mkern-13.5mu \lambda}
\def\ulambdabar{\,\lower 1.2ex\hbox{$\sim$}\mkern-13.5mu \lambdabar}
\def\ulambdabarzero{\,\lower 1.2ex\hbox{$\sim$}\mkern-13.5mu \lambdabar^\zero}
\def\ulambdabarnew{\,\lower 1.2ex\hbox{$\sim$}\mkern-13.5mu \lambdabar^\new}
\def\D{{\cal D}}
\def\M{{\cal M}}
\def\N{{\cal N}}
\def\Dslash{\,\,{\raise.15ex\hbox{/}\mkern-12mu \D}}
\def\Dbarslash{\,\,{\raise.15ex\hbox{/}\mkern-12mu {\bar\D}}}
\def\delslash{\,\,{\raise.15ex\hbox{/}\mkern-9mu \partial}}
\def\delbarslash{\,\,{\raise.15ex\hbox{/}\mkern-9mu {\bar\partial}}}
\def\L{{\cal L}}
\def\hf{{\textstyle{1\over2}}}
\def\quarter{{\textstyle{1\over4}}}
\def\eighth{{\textstyle{1\over8}}}
\def\fourth{\quarter}

\def\xibar{\bar\xi}

\def\uvcl{{\uv}^\cl}
\def\uAcl{\,\lower 1.2ex\hbox{$\sim$}\mkern-13.5mu A^{}_{\cl}}
\def\uAbarcl{\,\lower 1.2ex\hbox{$\sim$}\mkern-13.5mu A_{\cl}^\dagger}

\def\ASDzero{{{\scriptscriptstyle\rm ASD}\zero}}
\def\SDzero{{{\scriptscriptstyle\rm SD}\zero}}
\def\SD{{\scriptscriptstyle\rm SD}}

\def\three{{\scriptscriptstyle(3)}}
\def\dagthree{{\dagger\scriptscriptstyle(3)}}

\def\xione{\xi_1}
\def\xionebar{\bar\xi_1}
\def\xitwo{\xi_2}
\def\xitwobar{\bar\xi_2}

\def\Ltwo{\L_{\sst SU(2)}}
\def\Leff{\L_{\rm eff}}
\def\Laux{\L_{\rm aux}}
\newsec{Introduction}

\subsec{The instanton series in $N=2$ \susic\ gauge theory}

The last two years have seen remarkable progress in the study of $N=2$
supersymmetric Yang-Mills theories. This progress was initiated by the work
of Seiberg and Witten \SWone, who determined the exact low-energy effective
Lagrangian for the gauge group $SU(2)$. 
 Their work, which relies on a novel version
of Montonen-Olive duality \Olive, has subsequently been generalized to include
various matter couplings \SWtwo\ 
and larger gauge groups \refs{\KLTYone\AFone\KLTtwo\DS-\BIone}.
In $N=2$ \susic\ Yang-Mills theory, 
the low-energy effective Lagrangian is determined in terms of
a single object: the prepotential ${\cal F}$  \refs{\Gates,\Sone}.  The
prepotential is a holomorphic function of 
the $N=2$ chiral superfield \GSW\ $\Nequalstwo$
and the dynamically generated scale of the theory $\Lambda$. By
determining its behavior in the vicinity of its singularities, Seiberg
and Witten were able to reconstruct it exactly:
\eqn\FSWdef{\F(\Nequalstwo)\ \equiv
\ {\cal F}_{\rm pert}(\Nequalstwo)\,+\,{\cal F}_{\rm inst}(\Nequalstwo)
\ =\
{i\over2\pi}
\Nequalstwo^2\,\log{2\Nequalstwo^2\over e^3\Lambda^2}\ -\ {i\over\pi}
\sum_{n=1}^\infty
{\cal F}_n\,\left({\Lambda\over\Nequalstwo}\right)^{4n}\Nequalstwo^2\ ,}
where the inverse powers of $\Nequalstwo$ 
are understood in the sense of a Taylor  expansion about the vacuum
expectation value (vev), $\vhiggs$.
That the expansion has this general form had been known for some time 
\Sone$\,$; the $n$th term in the series is renormalization group invariant 
and has precisely the right transformation
properties under the anomalous $U(1)$ symmetry to be identified with 
an $n\hbox{-}$instanton effect. The new
information in the Seiberg-Witten solution is the precise numerical
value of each of the coefficients ${\cal F}_{n}$; in
particular,\foot{Numerical values of ${\cal F}_{n}$ depend 
on the prescription for $\Lambda$
which is fixed below as per Ref.~\FPone. In our conventions, 
the ${\cal F}_{n}$ are those of Ref.~\KLTtwo\ times a factor of $2^{6n-2}\,$.}
${\cal F}_1=1/2$  and ${\cal F}_2=5/16\,,$ with the
higher ${\cal F}_n$'s being easily determined by a recursion relation 
\MAtone.   This constitutes a set 
of highly non-trivial predictions for {\it all} 
multi-instanton contributions to the low-energy physics in this theory.

In principle, it should be possible to calculate the instanton
series without appealing to duality,
by directly evaluating the saddle-point contributions to
the path integral in the semiclassical limit. Previously this has only
been accomplished in the 1-instanton sector \refs{\FPone,\ISone}. 
In this paper, we examine the role of the complete set of multi-instantons,
constructed long ago by  Atiyah, Drinfeld, Hitchin and
Manin (ADHM) \refs{\ADHMone,\ADHMtwo}. At a purely formal level, we succeed in
recasting the $\{{\cal F}_n\}$ as integrals over the $8n$-parameter
families of ADHM collective 
coordinates, together with their superpartners.
 For $n=1$ and $n=2$ we are able to perform these integrations explicitly, and
confirm the predictions of Seiberg and Witten.   In Ref.~\dkmtwo\ we extend
our analysis to include $N=2$ matter hypermultiplets, paralleling
Ref.~\SWtwo.

Our motivation is twofold.  First, our calculation
serves as  an independent check on the proposed exact solution and therefore   
on the realization of electric-magnetic duality 
on which it relies. A second motivation concerns the old questions 
about the role of instantons in the strongly-coupled vacuum of QCD. 
The availability of exact information about multi-instanton effects 
in this theory makes it a promising 
theoretical laboratory for investigating this issue. Indeed
 several important simplifications occur for
instanton physics in this model. In particular, the fact that the
${\cal F}_{n}$ are independent of the coupling implies a
powerful non-renormalization theorem for  perturbation theory in
 instanton backgrounds. And in addition, while usually an expansion such
as Eq.~\FSWdef\  makes sense only in weak coupling ($\vhiggs\gg\Lambda$),
the exact solution may be analytically continued
all the way into the strong-coupling regime, up to 
the singular points  at which the theory admits a dual description in terms
of massless monopoles. This 
suggests the interesting possibility that the strongly-coupled vacuum 
can be described as a non-dilute gas of 
instantons which undergoes a phase transition near the singular point.

The general approach we adopt, and extend, herein was originally
developed by Affleck, Dine and Seiberg for $N=1$ models \ADSone, and 
adapted to the $N=2$ case by Seiberg \Sone. In this approach, the
long-distance physics is studied by focusing on certain
chirality-violating antifermion Green's functions,
$\langle\psibar(x_1)\,\psibar(x_2)\rangle$ in the $N=1$ theories
or $\langle\psibar(x_1)\,\psibar(x_2)\,\lambdabar(x_3)
\lambdabar(x_4)\rangle$ in the $N=2$ theories as dictated by the
anomaly structure. These Green's functions are then saturated
in saddle-point approximation by their classical values,
$\psibar\rightarrow\psibar_\cl$, etc., obtained by solving the
Dirac equation in the classical instanton background.

Strictly speaking, when the gauge group is  spontaneously broken,
the instantons  are no longer exact solutions of
the equations of motion \Derrick.
 We follow Affleck et al.~\refs{\Aone,\ADSone} and work instead with   the
\it constrained \rm instanton (see Sec.~3). 
Despite its approximate nature,  this 
field configuration has  universal properties both at
short and long distances which yields unambiguous answers
for the corresponding asymptotics of the
 correlation functions. At lowest order in $g^2$,
the short-distance constrained instanton simply coincides with the
exact self-dual solution of the unbroken theory. Furthermore,
in  supersymmetric models, the leading-order short-distance fermions and
scalars assume their classical zero-mode values in this self-dual background. 
In contrast, 
at large distances, the fields satisfy linearized equations; nevertheless
the overall amplitudes of the long-distance fields
are determined by the  short-distance behavior \refs{\Aone,\ADSone}.

When one generalizes to $n>1$ there
is an additional set of issues to be confronted.
The saddle-point contribution of ADHM multi-instantons to the 
functional integral for pure bosonic Yang-Mills
 theory has been studied extensively
with inconclusive results.\foot{See Ref.~\Oone\ for a review of
progress made in this direction as well as a useful set of
references.}
 This program was never completed for several reasons. 
First, it was not possible to calculate the functional
determinant for small fluctuations around the general ADHM solution
which  provides a vital prefactor for the  
saddle-point exponential. Second, the calculations were plagued with the
usual infrared divergent integrals over instanton sizes which are ubiquitous in
theories with classical scale invariance. 
Finally the parameters which appear in the general ADHM
solution are not independent but obey non-linear constraints. These
constraints have only been solved explicitly for $n\leq 3$ \CWSone. 
Fortunately, in the supersymmetric  theory at hand, 
the first two difficulties are simply not
present. The non-zero eigenvalues in the functional determinants 
for small fluctuations cancel 
precisely between bosonic and fermionic degrees of freedom.
In fact this is just
the manifestation at one loop of the non-renormalization
theorem mentioned above.     
The second problem is eliminated because the vev acts
as an infrared cut-off on the contribution of large instantons. 
The third problem is harder to avoid but
there are preliminary indications that supersymmetry may provide
simplifications in this regard also.
Certainly it would be gratifying if knowledge of the specific
instanton series \FSWdef, in this one particular model,
 were to lead to fundamental progress
in the general theory of multi-instantons.

\subsec{The plan of this paper}

This paper is organized as follows. 
In Sec.~2, we review both the microscopic  $SU(2)$ gauge theory,
 and also, the effective long-distance
  $U(1)$ theory valid for length scales $\gg\,1/g\vhiggs$.
In Sec.~3 we discuss the coupled Euler-Lagrange equations at
both distance scales, as well as the ``patching condition'' 
which strictly relates the tail of the constrained instanton to its 
core \refs{\Aone,\ADSone}. 

The heart of the paper is the study of ADHM multi-instantons,
Secs.~6-8. However, as essential groundwork, we first present
a detailed study of the single superinstanton sector in Secs.~4-5. 
This subject has been studied extensively 
\refs{\ADSone,\NSVZone\FSone\FUone-\Amati}; in particular
Finnell and Pouliot \FPone\ have applied the methods of
\ADSone\ to the $N=2$ theory, and confirmed the 1$\hbox{-}$instanton
coefficient ${\cal F}_1=1/2$ in Eq.~\FSWdef. Nevertheless there
are two reasons for revisiting this subject in a careful way.
The first is curiosity about the instanton methods themselves;
we would like to see how to reproduce the full 1-instanton
sector of Eq.~\FSWdef, not just the 4-fermi vertex. Specifically, this means
understanding the fermion bilinear contributions to the 
classical bose fields. The detailed structure of the superinstanton
is the subject of Sec.~4. We then recover, in Sec.~5, the 1-instanton
contribution to \it all \rm the effective Seiberg-Witten vertices that
can be probed by saturating $m$-point functions with $m$ insertions
of long-distance classical fields (e.g., the anomalous magnetic moment
vertex).
Ultimately, 
however, there is a much more important reason to revisit the 1-instanton
sector: as we shall make clear,
detailed knowledge of these fermion-bilinear contributions
provides an {\it explicit roadmap}\foot{Technically, 
the reason that the problem
is inherently much more challenging for $n>1$ can be traced to
 the following: In the
1-instanton sector, the full complement of fermion zero modes can
be generated by applying Lagrangian symmetries to the BPST
instanton \BPST, specifically the generators of the superconformal group
\refs{\NSVZone-\FUone} (Appendix A). 
But for $n>1$ these only sweep out a fixed
subset of the $8n$-dimensional space of fermionic
collective coordinates, so that the methods of Refs.~\refs{\NSVZone-\FUone}
are insufficient.} for how to solve the analogous equations
for $n>1.$  

 Section 6 is a 
self-contained steepest-descent introduction to the ADHM construction
of multi-instantons. In Sec.~7 we apply the lessons learned in 
Secs.~4-5 to the more challenging problem of solving the coupled
boson-fermion equations of motion in the general ADHM background.
Based on our explicit
1-instanton expressions, we are able to intuit the correct solutions,
and verify our guesswork \it a posteriori\rm. The principal result
of this Section, and of the paper as a whole, is the explicit construction
of the adjoint Higgs in the exact ADHM multi-instanton background,
and with it, the exact classical interaction between $n$ instantons,
both in the bosonic and in the supersymmetric 
cases (Eq.~(7.32)). As expected on
general grounds from Eq.~\FSWdef, 
this interaction lifts all but four of the $8n$ fermion zero  modes.

Finally, in Sec.~8 we specialize to $n=2.$  
 The main technical hurdle in this Section is
the construction of the 2-instanton measure.
In this task we rely heavily on some remarkable results due
 to Osborn \Oone\ and to Corrigan and collaborators
\refs{\CGTone,\CGone}, which are reviewed as needed. We also supply some
essential new ingredients: principally, the calculation of the fermion
zero mode determinant (Appendices B and C), and the pinpointing of the residual
discrete symmetry
 group that  must be modded out in order to identify the physical moduli
space of inequivalent field configurations, and thus properly to
normalize the measure (Appendix D). With a convenient
resolution of ADHM constraints the 28$\hbox{-}$fold
integration over the \susic\ moduli proceeds in a straightforward manner,
and confirms that ${\cal F}_2=5/16.$

\newsec{Microscopic vs.~Effective Lagrangians}

In this Section we review both the microscopic  nonabelian  theory
where the instantons live, and also, the effective long-distance
abelian theory that the power-law ``tail'' of the instanton is 
supposed to reproduce.  
The particle content of pure $N=2$ \susic\ $SU(2)$ gauge theory consists, in
$N=1$ language, of a gauge multiplet $(\uv_m,\ulambda\,,\uD\,)$ coupled
to a complex chiral matter multiplet $(\uA,\upsi,\uF\,)$ which transforms in
the adjoint representation of the gauge group. 
Here $\uv_m$ is the gauge field, $\uA$ is the Higgs field, Weyl fermions 
$\ulambda$ and $\upsi$ are the gaugino and Higgsino,
 while $\uD$ and $\uF$ are auxiliary fields. 
In Wess and Bagger  notation \wessbagger, the component
Lagrangian reads\foot{Note on conventions: we use undertwiddling 
as a shorthand for $SU(2)$ matrix notation; thus
${\uX}=\sum_{a=1,2,3}\,X^a\tau^a/2$, where $\tau^a$ are Pauli matrices.
 Letters from the beginning of
the alphabet are adjoint $SU(2)$ indices running over $1,2,3$ whereas
letters from the middle of the alphabet run over $0,1,2,3$ (or in
Euclidean space $1,2,3,4$).  Also
$\uv_{mn}\ =\ \partial_m\uv_n-\partial_n\uv_m-ig\,[\,\uv_m,\uv_n\,]\,,$
$\Dslash_{\alpha\dot\alpha}\ =\ \D_m\sigma^m_{\alpha\dot\alpha}$, and
$\Dbarslash^{\dot\alpha\alpha}\ =\ \D^m\sigmabar_m^{\dot\alpha\alpha}$,
where $\,\D_m{\uX}\ =\ \partial_m{\uX}-ig\,[\,\uv_m,{\uX}\,]$.
Wess  and Bagger conventions are used throughout \wessbagger:
 $\chi \zeta \ = \ \chi^{\alpha} \zeta_{\alpha}$, 
$ \bar\chi \bar\zeta \ = \ \bar\chi_{\dot\alpha} \bar\zeta^{\dot\alpha}$,
$\chi\sigma^{mn}\zeta\ =\ \chi^\alpha\sigma^{mn}{}_\alpha{}^\beta\zeta_\beta,$
$\bar\chi\sigmabar^{mn}\bar\zeta\ =\ \bar\chi_\dalpha
\sigmabar^{mn\,\dalpha}{}_\dbeta\bar\zeta^{\,\dbeta}\,.$ The metric
is $\eta_{mn}=\hbox{diag}(-1,1,1,1)$. Throughout this paper we work
in Minkowski space even when dealing with instantons; analytic continuation
to Euclidean space poses no problems. Through Sec.~5 we will usually
keep factors of $g$ explicit, unlike Seiberg and Witten who set
$g=1$; thus for instance their condition for weak coupling, $\vhiggs\gg
\Lambda,$ becomes for us $M_W\gg\Lambda.$
}
%
\eqn\Ltwodef{\Ltwo\ =\ \L_{\rm gauge}+\L_{\rm chiral}}
where
\eqn\Lgaugedef{\L_{\rm gauge}=\trtwo\Big\{-\hf\uv_{mn}\uv^{mn}
-i\ulambdabar\Dbarslash\ulambda
-i\ulambda\Dslash\ulambdabar
+\uD^2\Big\}}
and
\eqn\Lmatterdef{
\eqalign{\L_{\rm chiral}\ =\
\trtwo\Big\{&-2\D_m\uAbar\D^m\uA
-i\upsibar\Dbarslash\upsi
-i\upsi\Dslash\upsibar
+2\uFbar\uF
\cr&-2g\uD\,[\,\uA\,,\uAbar\,]\,+2\sqrtwo 
gi\Big(\,[\,\uAbar,\upsi\,]\,\ulambda
+\ulambdabar\,[\,\uA\,,\upsibar\,]\,\Big)\Big\}\ .}}
In addition to superconformal invariance (see Appendix A for
a review), $\Ltwo$ is
classically invariant under so-called $SU(2)_\R$ rotations of 
 $\ulambda$ and  $\upsi$:
\eqn\SUtwoR{\pmatrix{\ulambda\cr\upsi}\ \longrightarrow\
{\Mr}\cdot
\pmatrix{\ulambda\cr\upsi}\ ,\quad
\Big(\ulambdabar\ ,\ \upsibar\,\Big)\ \longrightarrow\
\Big(\ulambdabar\ ,\ \upsibar\,\Big)\cdot\Mr^{-1}\ ,
\quad\Mr\,\in\,SU(2)\ .}
The bosons are $SU(2)_\R$ singlets. $N=1$ invariance together
with $SU(2)_\R$ invariance guarantees $N=2$ invariance, a labor-saving
observation which we exploit below.

Eliminating the auxiliary field $\uD$ from $\Ltwo$ produces a scalar potential
proportional to $\trtwo\,[\uA,\uAbar]^2$. The classical
vacua of the theory therefore
consist of all constant fields $\uA$ and $\uAbar$ satisfying
$\uA\propto\uAbar.$ With a gauge transformation we can always
align the vev along a specific direction,
for instance $\uA=\vhiggs\tau^3/2$ and
$\uAbar=\vbarhiggs\tau^3/2$; the gauge inequivalent
vacua are then labeled by the arbitrary complex number $\vhiggs$.
 Importantly, since $N=2$ \susy\ protects
against the generation of a superpotential, this degeneracy of vacua
persists at the quantum level as well \Sone.

For nonzero $\vhiggs$ the gauge group $SU(2)$ spontaneously breaks down
to $U(1)$. The components of the fields that are aligned with the vev
remain massless, and are neutral under the unbroken $U(1)$, whereas
the remaining charged components acquire a mass $M_W=\sqrtwo g|\vhiggs|.$
For length scales $x\gg1/M_W$ the massless modes can be described by an
effective Lagrangian $\Leff$, constructed from a single $N=2$ superfield
$\Nequalstwo,$ or, in more familiar language, from an $N=1$ photon superfield
$W^\alpha=(v_m,\lambda,\rmD)$ and chiral superfield $\Phi=(A,\psi,F).$
Underlying $N=2$ \susy\ forces $\Leff$ to have the following
form:
\eqn\Leffdef{\Leff\ =\ {1\over4\pi}\,\Im\left[\int d^4\theta\F'(\Phi)
\Phibar+\int d^2\theta\,\hf\F''(\Phi)W^\alpha W_\alpha\,\right]\ +
\ \cdots\ ,}
where the prepotential $\F$ was defined in Eq.~\FSWdef, and 
the dots represent terms of higher order in ``chiral perturbation
theory.'' 
In this paper we will be focusing on the region of moduli space $M_W\gg
\Lambda$ where the theory is weakly coupled due to asymptotic freedom,
and instantons may be studied in the semiclassical approximation.
Nevertheless we ought to mention an especially surprising feature of
the exact Seiberg-Witten solution in the opposite regime, namely that
the classical symmetry-restoring point $\vhiggs=0$ of the effective theory
is actually absent from the quantum moduli space.

As with the microscopic theory it will be useful  to write 
out $\Leff$ in components:
\eqn\Leffcomp{\eqalign{\Leff\ =\ {1\over4\pi}\,\Im\Big[
&-\F''(A)\Big(\partial_mA^\dagger\partial^mA+i\psi\delslash\psibar
+i\lambda\delslash\lambdabar+\hf(\vsd_{mn})^2\,\Big)
\cr&+{\textstyle{1\over\sqrtwo}}\F'''(A)\lambda\sigma^{mn}\psi v_{mn}+\quarter
\F''''(A)\psi^2\lambda^2\,\Big]\ +\ \Laux\ .}}
Here $\Laux$ assembles all the dependence on the auxiliary fields
$F,$ $F^\dagger$ and $\rmD$:
\eqn\Lauxdef{\Laux\ =\ 
{1\over4\pi}\,\Im\left[\F''(A)\big(F^\dagger F+\hf \rmD^2\big)
-\hf \F'''(A)\big(F^\dagger(\psi^2+\lambda^2)-i\sqrtwo \rmD\psi\lambda\big)\,
\right]\ .} 
The quantity $\vsd_{mn}$ in \Leffcomp\ denotes the self-dual part of the field 
strength\foot{In Minkowski space the self-dual and anti-self-dual
components of $v_{mn}$ are projected out using
$\vsd_{mn}\ =\ \quarter\big(\eta_{mk}\eta_{nl}-\eta_{ml}
\eta_{nk}+i\epsilon_{mnkl}\big)v^{kl}$ and $\vasd_{mn}=(\vsd_{mn})^*$,
where $\epsilon_{0123}=-\epsilon^{0123}=-1$. Also, since
$\sigma^{mn}=\fourth\,\sigma^{[\,m}\sigmabar^{n\,]}$ and
$\sigmabar^{mn}=\fourth\,\sigmabar^{[\,m}\sigma^{n\,]}$ are
self-dual and anti-self-dual, respectively, it follows that
$\sigma^{mn\ \beta}_{\phantom{mn}\alpha}v_{mn}
=\sigma^{mn\ \beta}_{\phantom{mn}\alpha}\vsd_{mn}$
and
$\sigmabar^{mn\dot\alpha}_{\phantom{mn\alpha}\dot\beta}v_{mn}
=\sigmabar^{mn\dot\alpha}_{\phantom{mn\alpha}\dot\beta}\vasd_{mn}$.
}
so that $\big(\vsd_{mn}\big)^2$ comprises both $(v_{mn})^2$ and
$i v_{mn}v_{kl}\epsilon^{mnkl}$ terms. For these two terms  to have
their customary couplings, 
 $\F''$ at the vev scale, or indeed at any scale $\mu,$ 
must be identified with the complexified coupling $\tau (\mu)$:  
\eqn\tauequals{\F''(\mu) 
\ = \ \tau (\mu) \ = \ {4 \pi i \over g^2 (\mu)} + 
   {\vartheta (\mu) \over 2 \pi} \ ,}
where $\vartheta$ is the effective theta-parameter.
Note that in this paper we are using the notation of Ref.~\SWone\ for
the complexified coupling \tauequals. Trivial rescaling connects it
with the notation of Ref.~\SWtwo: multiply the right-hand side of
Eq.~\tauequals\ by 2, and divide by 2 the right-hand sides of
Eqs.~\Leffdef-\Lauxdef\ and \FSWdef.

We make several comments:

\bf1\rm. The correspondence between the RG-invariant 
dynamical mass scale $\Lambda$
in the effective $U(1)$ theory, and the running coupling $g(\mu)$
in the microscopic $SU(2)$ theory, has been carefully examined by Finnell
and Pouliot \FPone, using 
matching to perturbation theory in the weak coupling regime.
In the Pauli-Villars regularization scheme
 which is the natural scheme for doing instanton calculations,
the result of \FPone\ is simply
\eqn\FPdef{\Lambda^4\ =\ 
\Lambda_{\sst\rm PV}^4 \ \equiv\ \mu^4\,e^{-8\pi^2/g_{\sst\rm PV}^2(\mu)}\ .}
A different choice of $\Lambda$ would imply an altered prescription for the 
$\{{\cal F}_n\}$. 
{}From now on the Pauli-Villars scheme will be always assumed 
and in what follows we suppress the PV subscript.

\bf2\rm. As always in effective theories, the relationship between the
effective fields $\{A, \psi, \dots\}$ and the  microscopic fields
$\{\uA, \upsi, \dots\}$ is in no way unique, and may be quite complicated.
But for long-distance physics ($x\gg 1/M_W$) 
we can simply equate the effective fields with the surviving massless
components; thus $A=\hf g\,\trtwo\tau^3\uA,$ 
$\psi=\hf g\,\trtwo\tau^3\upsi,$ etc.,
assuming the vev points in the $\tau^3$ direction. The factor of $g$
compensates for the overall normalization of $\Leff$  by $1/g^2$
as opposed to $\Ltwo$.

\bf3\rm. Integrating out the auxiliary fields from Eq.~\Lauxdef\ gives
new contributions to the effective
vertices, for instance to the $\psi^2\lambda^2$
operator. However these are down by a relative factor of
$\big(\Im\,\F''(A)\big)^{-1}\sim g^2$ compared with existing vertices,
term by term in the instanton expansion. For simplicity we ignore
$\Laux$ in the following, and choose to concentrate, at any given order in
the instanton expansion, on the leading contributions in $g^2.$
In principle, the contribution of $\Laux$ can be recaptured 
with instanton methods by analyzing
Feynman graphs in instanton backgrounds.

\newsec{Defining Equations for the Constrained Superinstanton}

Throughout this paper, we will use the term ``superinstanton'' in a loose way,
simply to denote a nontrivial finite-action (exact or approximate)
solution to the
coupled \susic\ Euler-Lagrange equations of the theory.\foot{In particular,
we are making no claim about the exchangeability, at \it all \rm
length scales simultaneously, 
of an active \susy\ transformation on the fields for a passive
transformation on the collective coordinates, which property is at the
heart of Novikov et al.'s use of the term \NSVZone.}
For the model at hand, these equations read, in Minkowski space:

\eqna\eulergauge
$$\eqalignno{
\D^m\uv_{mn}\ &=\ -ig\Big(\,[\,\uA\,,\D_n\uAbar\,]\,+\,[\,\uAbar,\D_n\uA\,]\,
\Big)
\cr&\quad\
-g\Big(\ulambda\sigma_n\ulambdabar+\ulambdabar\barsigma_n\ulambda
+\upsi\sigma_n\upsibar+\upsibar\barsigma_n\upsi\Big) &\eulergauge a
\cr
\Dbarslash\ulambda\ &=\ \sqrtwo g\,[\,\uA\,,\upsibar\,]\,&\eulergauge b
\cr
\Dslash\ulambdabar\ &=\ \sqrtwo g\,[\,\uAbar,\upsi\,]\,&\eulergauge c
\cr
\uD\ &=\ g\,[\,\uA\,,\uAbar\,]\,&\eulergauge d
}$$
for the $N=1$ gauge multiplet;
\eqna\eulermatter
$$\eqalignno{
\D^2\uA\ &=\ \sqrtwo ig\,[\,\ulambda\,,\upsi\,]\,+g\,[\,\uD,\uA\,]\,
&\eulermatter a
\cr
\Dbarslash\upsi\ &=\ \sqrtwo g\,[\,\ulambdabar\,,\uA\,]\,
&\eulermatter b\cr
\uF\ &=\ 0&\eulermatter c}$$
for the $N=1$ chiral multiplet; and
\eqna\eulerantimatter
$$\eqalignno{
\D^2\uAbar\ &=\ \sqrtwo
 ig\,[\,\ulambdabar\,,\upsibar\,]\,-g\,[\,\uD,\uAbar\,]\,&\eulerantimatter a
\cr
\Dslash\upsibar\ &=\ \sqrtwo g\,[\,\ulambda\,,\uAbar\,]\,&\eulerantimatter b
\cr
\uFbar\ &=\ 0&\eulerantimatter c}$$
for the $N=1$ antichiral multiplet.

There are actually two philosophically distinct, but mathematically
equivalent, ways of viewing the fermion components in these equations.
In one approach, the classical background configuration is always
purely bosonic, and the fermions are treated as a particular set
of fluctuations. This is
the viewpoint that one finds in 't Hooft's original paper \tHooft,
and which is implicit in the language used by Affleck, Dine and
Seiberg \ADSone. In contrast, in the formalism of \novetal~\NSVZone,
the fermion  modes naturally appear as the superpartners of the classical gauge
and Higgs configurations (as we shall see below). 
In this approach the fermions acquire
a geometric meaning, and are thought of as facets  of the classical
solution. This is the viewpoint, and the language, we adopt herein.

Returning to these equations, we recall that
for nonzero vev  a nontrivial solution cannot exist, thanks to
Derrick's theorem \Derrick: 
for any putative solution one can lower the action further
simply by shrinking the configuration. One famous fix to this problem, due to
Affleck \Aone, is as follows. A
 new operator, or ``Affleck constraint,'' is introduced into the action
by means of a Faddeev-Popov
 insertion of unity. If this operator is of suitably high dimension,
Derrick's theorem is avoided,
 and the instanton stabilizes at a fixed scale size $\rho.$
The integration over the 
Faddeev-Popov Lagrange multiplier can then be traded off for the
integration over $\rho.$ 
The now-stable solutions are known as constrained instantons.

Of course, the detailed 
shape of the constrained (super)instanton depends in a complicated
way on one's choice of constraint. But  certain important features remain
constraint independent, namely:

\bf1\rm. The short-distance regime, $x\ll1/M_W.$ In this regime 
Eqs.~\eulergauge{}-\eulerantimatter{}
can be solved perturbatively in $g^2\rho^2|\vhiggs|^2\,$;
since ultimately the integration over scale size is dominated by $\rho\sim 
|\vhiggs|^{-1}$
this is tantamount to 
perturbation theory in $g^2.$ As the constraints do not enter
into these equations until  some high order, the first few
terms in this expansion are
 robust. For $\rho\ll x\ll1/M_W$ the various  fields
fall off as powers of $\rho^2/x^2$; again the first few terms in the expansion
in $\rho^2/x^2$ are constraint independent. Note that none of these
inequalities conflict in any way with the requirement for
semiclassical physics, $M_W\gg\Lambda.$

\bf2\rm. The long-distance
 regime, $x\gg1/M_W.$ The long-distance ``tail'' of the instanton
reflects the Higgs mechanism.
 In the model at hand, the superinstanton components
perpendicular to the vev decay
 as $\exp(-M_W|x|).$ In contrast, the components parallel
to the vev fall off merely as
 powers of $\rho^2/x^2.$  They are constrained to obey the $U(1)$
reductions of Eqs.~\eulergauge{}-\eulerantimatter{}, namely
\eqn\longeom{\eqalign{0 &= \partial^m\,v_{mn}^\three =
\delbarslash\,\lambda^\three =
\delslash\,\lambdabar^\three = \rmD^\three \cr&=
\partial^2\,A^\three = \delbarslash\,\psi^\three
= F^\three \cr
&= \partial^2\,A^\dagthree = \delslash\,\psibar^\three
= F^\dagthree \ ,}}
again assuming that the vev is aligned in the $\tau^3$ direction.
By themselves, these equations allow a distressingly broad range of
possible behaviors for the instanton tail; for example, the long-distance
Higgs can decay as $\vhiggs\,+\,\cld/(x-x_0)^2$ for any constant
$\cld.$ This is a hallmark of linear equations.
Fortunately, Eqs.~\longeom\ are supplemented by a
\it patching condition \rm which ties $\cld$ to the analogous
short-distance constant $\csd$ derived in the regime
$\rho\ll x\ll1/M_W.$ Briefly, the patching condition states
\eqn\patchdef{\cld\ =\ \csd\cdot\big(1\,+\,O(g^2)\,\big)\ ,}
with the specific form of the
Affleck constraints only affecting these $O(g^2)$ corrections.

It is precisely because of this patching condition that it is important
 to study the short-distance properties of the instanton---even if, as
here, one is ultimately interested in the tail.\foot{As 
opposed to the short- and 
long-distance regimes, very little can be said about the
shape of the instanton at 
length scales $x\sim1/M_W.$ This is the domain where the
short- and long-distance behaviors are patched together, and the effect of the
constraints is unsuppressed. While it underlies the work of Affleck,
Dine and Seiberg, and much of the literature on high-energy baryon number
violation, so far as we know the patching condition is still at the level
of a ``folk theorem'' about nonlinear differential equations; a rigorous
proof would start by recasting these as integral equations.}
Accordingly, in the following Section we give an unusually careful analysis of
the short-distance superinstanton, and are twice rewarded
for our efforts. First, this will enable us to rederive
\it all \rm of the pieces of $\L_\eff$ that can be
obtained by saturating $n\hbox{-}$point functions with $n$ insertions of
single instantons.
And second,  our explicit 1-instanton results will allow  us later to intuit
the general solutions to Euler-Lagrange equations in arbitrary
multi-instanton backgrounds.

\newsec{The Single Short-Distance Super-Instanton}

\subsec{Overview of calculation}

In this Section and the next we
specialize to the one-instanton sector.\foot{Those 
readers primarily interested in what we have
to say about multi-instantons may skip directly to Sec.~6.}
 In this case the 
short-distance superinstanton is most easily constructed with the
``sweeping out'' technique originated
by Novikov \etal~\NSVZone, and extended by Fuchs and 
Schmidt \refs{\FSone-\FUone}. By design, this
technique covers the solution space of the coupled Euler-Lagrange
equations, to any desired order in $g^2\rho^2|\vhiggs|^2.$
Specifically, starting with a ``reference'' configuration $\Psi^\zero$
which solves Eqs.~\eulergauge{}-\eulerantimatter{} at a given
order in perturbation theory, the superinstantons are then the family of
configurations generated by
\eqn\family{e^{\xi(x)Q}\times e^{\Thetabar\Qbar}\times \Psi^\zero\ .}
$\Psi^\zero$ stands for the initial values for all the bosons and
fermions in the model.
As explained in Appendix A, the product $\xi(x)Q$ encapsulates
 both  $Q^\alpha$ and $\Sbar_{\dot\alpha}$, where $Q$ is the
$N=1$ \susy\ generator, and $\Sbar$ is the fermionic superpartner
of the special conformal generator $K_\mu$.
The action of $\Qbar,$ $Q$ and $\Sbar$ on all the component
fields in $\Psi^\zero$ is specified in Eqs.~(\vmntrans)-(\sconfdef).

The calculation that follows has the following features:

\bf1\rm. Since the construction \family\ is only manifestly $N=1$
invariant,\foot{While, in principle, an $N=2$ invariant sweeping-out
procedure would be more aesthetic for the particular problem at hand, 
in practice it seems much harder to carry out, the primary reason being that
$N=2$ invariant Grassmannian exponentials such as $\exp(\bar\Theta^\one
\Qbar^\one+\bar\Theta^\two\Qbar^\two)$ terminate at quartic
rather than quadratic order. This technical difficulty, together with
the desirability of making contact with the work of Novikov \etal\ and
Fuchs and Schmidt, leads us to favor the $N=1$ formalism.} it is necessary
also to check $SU(2)_\R$ invariance, Eq.~\SUtwoR. We shall find the
following: at leading order, the $\upsi$ and $\ulambda$
components of the superinstanton appear in an $SU(2)_\R$ symmetric
way, but $\upsibar$ and $\ulambdabar$ do not; at the next-leading
order $SU(2)_\R$ invariance of the antifermions is recaptured, but
now the fermions fail to enter symmetrically by a small amount;
and so forth, order by order in $g^2\rho^2|\vhiggs|^2$. Only at
infinite order is exact $SU(2)_\R$ symmetry and hence
$N=2$ supersymmetry manifest,
assuming  that the Affleck constraints themselves respect
$N=2$ \susy\ (which in principle they need not).

\bf2\rm. This ``leapfrog'' pattern characterizes the Euler-Lagrange
equations \eulergauge{}-\eulerantimatter{} as well. At leading order
the $N=1$ matter equations \eulermatter{}-\eulerantimatter{} are exact whereas
the gauge equations \eulergauge{}\ are approximate; at next-leading
order the latter become exact while the former are off by
a small amount; and so forth.

In practice we content ourselves with a next-leading order calculation,
where nontrivial checks on the proposed Seiberg-Witten solution can first
be made. Our explicit next-leading-order results, as well as our
formulae for the components of the antichiral superfields, are beyond
the existing literature.

\subsec{Leading-order calculation}

Following Fuchs and Schmidt \FSone, we start with a reference
configuration $\Psi^\zero$ which is \it almost \rm purely bosonic.
Specifically, we take for the gauge field $\uv^\zero_m=\uv^\cl_m$
where $\uv^\cl_m$ is a BPST instanton \BPST\ having some particular
position $x_0,$ scale size $\rho,$ and iso-orientation $R_{ab}$. We
take for the Higgs $\uAzero=\uAcl$ and $\uAbarzero=\uAbarcl$, where
 $\uAcl$ satisfies the massless Klein
Gordon equation in the BPST background, $\D^2\uAcl=0$,
while approaching the vev $\vhiggs^a\tau^a/2$
at infinity. All other bosonic and fermionic components are initially
set to zero, \it except \rm for $\upsi$. As noted by Fuchs and 
Schmidt,\foot{This is a minor peculiarity of the $N=1$ approach in the
presence of an adjoint Higgs, which we could avoid by adopting from the outset
an $N=2$ sweeping-out procedure. Our choice of notation $\xi'$ reminds
us that this is really the collective coordinate for the second
supersymmetry; see Eq.~(\susytwodef$f$).}
this component needs to be loaded initially with a supersymmetric zero
mode, else none is generated by the sweeping-out procedure \family.
Thus we set 
\eqn\psizerodef{\upsizero\ =\ \xi'\sigma^{mn}\uv^\cl_{mn}\ ,}
 where $\xi'$ is a new Grassmannian collective coordinate.
This defines our starting point $\Psi^\zero.$

Sweeping out as per Eq.~\family\ is a two-step process.  First one
applies $\exp(\Thetabar\Qbar)$ to each component field, and refers
the final expressions to their prescribed values in $\Psi^\zero$ (see
Eq.~(\example) for an example). With $\Psi^\zero$ as specified above, and
remembering that $\sigmabar^{mn}\uv_{mn}^\cl=0$ by self-duality, we find
that the \it only \rm field that transforms is the Higgsino, which
picks up an admixture of a superconformal zero mode (see Appendix A):
$\upsi\rightarrow\upsizero-i\sqrtwo\Thetabar\Dbarslash\uAcl.$ 
Next, one repeats the procedure with $\exp(\xi(x)Q).$ The result of
this simple
exercise is the leading-order superinstanton with the following functional
 form for the gauge, chiral and antichiral multiplets, respectively:
\eqna\losuper
$$\eqalignno{\uv_m&=\uv^\cl_m\ ,\qquad \ulambda=
-\xi(x)\sigma^{mn}\uv^\cl_{mn}\ ,
\qquad  \ulambdabar=\uD=0\ ;&\losuper a
\cr
\uA&=\uAcl+\sqrtwo\xi'(x)\sigma^{mn}\xi(x)\uv^\cl_{mn}\ , \quad 
\upsi=\xi'(x)\sigma^{mn}\uv^\cl_{mn}\ ,\quad \uF=0 ;\ &\losuper b\cr
\uAbar&=\uAbarcl\ ,\qquad \upsibar=-i\sqrtwo
\big(\xi(x)\Dslash+2\etabar\big)\uAbarcl \ ,\qquad \uFbar=0\ .&\losuper c}$$
The reader can verify that Eqs.~\losuper{a,b} are equivalent, up to
a collective coordinate dependent gauge transformation, to the elegant
superfield expressions for $\uW^\alpha_\cl$  and $\uPhi_\cl$,
respectively, derived
by \Novetal~\NSVZone. For present purposes, however, Eq.~\losuper{}
has the advantage of commuting with gauge fixing. This will allow for
a clean physical interpretation for the quanta that propagate
to infinity $\grave{\rm a}$ la Affleck, Dine and Seiberg \ADSone.

In arriving at Eq.~\losuper{} we have used the equations of motion
\eulergauge{}-\eulerantimatter{} to eliminate all quadratic terms
in the expansions of the Grassmann exponentials. Furthermore, in order to
highlight the manifest $SU(2)_\R$ symmetry between $\ulambda$ and $\upsi,$
we have traded  $\Thetabar$ for a linearly
related collective coordinate $\etabar',$
defined in analogy to Eq.~(\xiofxdef), as follows:
\eqna\etabarprimedef
$$\eqalignno{\xi'(x)\ 
&=\ \xi' - (x^k_{}-x^k_0)\etabar'\sigmabar_k\ ,&\etabarprimedef a
\cr
\etabar'\ &=\ \textstyle{1\over8}i\sqrtwo\,
g\,\vhiggs^a \etabar^{\prime a}_{mn}\Thetabar\sigmabar^{mn}\ ,
&\etabarprimedef b\cr
\Thetabar\ &=\ i\sqrtwo\,
g^{-1}\,{\vbarhiggs^a\over|\vhiggs|^2}\,
\etabar^{\prime a}_{mn}\,\etabar'\sigmabar^{mn}\ .&\etabarprimedef c}$$
Here $\etabar^{\prime a}_{mn}$ is short for $R_{ab}\,\etabar^b_{mn}\,,$
with $\etabar^b_{mn}$ an 't Hooft symbol \tHooft. 
This definition allows us to translate back and forth between the two
distinct-looking representations of the superconformal zero mode
(see Appendix A), since
\eqn\sconfrewrite{i\sqrtwo \Thetabar\Dbarslash\uAcl\  =\ 
(x^{}_k-x^0_k)\etabar'
\sigmabar^k\sigma^{mn}\uv_{mn}^\cl}
 as the reader can verify in his favorite gauge (e.g., Eq.~(4.8) below).

As is clear from Eq.~\losuper{},
$SU(2)_R$ symmetry, originally
defined in Eq.~\SUtwoR\ as an active transformation
on the fermions,  can be recast as a passive transformation on
the Grassmannian collective coordinates:
\eqn\SUtwoRb{\pmatrix{\xi(x)\cr-\xi'(x)}\ \longrightarrow\
{\Mr}\cdot
\pmatrix{\xi(x)\cr-\xi'(x)}\ .}
Note that the fermion bilinear contribution to $\uA$, while
 breaking the symmetry between $\uA$ and $\uAbar,$ is an $SU(2)_\R$
singlet under \SUtwoRb, as it must be.

In contrast,
$SU(2)_\R$ is \it not \rm respected by the antifermions; indeed
$\ulambdabar$ is not even turned on at this order.
And  while Eqs.~\eulermatter{}-\eulerantimatter{}
are exactly satisfied by the leading-order superinstanton \losuper{},
 Eqs.~\eulergauge{}\ are not; instead, what is satisfied are the
homogeneous variants of Eqs.~\eulergauge{}, with the right-hand sides
set to zero. Both these drawbacks  are rectified
at next-leading order, as mentioned above, and confirmed in Sec.~4.4 below.

\subsec{Superinstanton action}

Knowledge of the leading-order short-distance superinstanton \losuper{}
suffices to construct the action, up to $g^2\rho^2|\vhiggs|^2$ corrections.
We shall do so in the way that generalizes most readily
to multi-instantons, by expressing the answer as a surface
integral.  At leading order the only non-vanishing terms are
\eqn\instaction{\eqalign{S_\inst\ &\equiv\ S_\cl+S_\higgs+S_\yuk
\cr&=\ \trtwo\int d^4x\,\Big(\ -\hf\,\uv_{mn}\uv^{mn}\ 
-\ 2\,\D_m\uAbar\D^m\uA\ +\
2\sqrtwo gi\,[\,\uAbar,\upsi\,]\,\ulambda\ \Big)
\cr&=\ {8\pi^2\over g^2}\ 
-\ 
2\,\trtwo\int d^3{\rm S}\,\uAbar\,\hat x_m\,\D^m \uA\ ,}}
where $\hat x_m=x_m/\sqrt{|x|^2}$ and S is the 3-sphere at 
infinity.\foot{Actually up to subleading corrections the radius of the
sphere should be taken to be $\gg\,\rho$ but $\ll\,1/M_W$ so that the
short-distance formulae \losuper{} are applicable.}
The last equality follows from an integration by parts together with
the Euler-Lagrange equation \eulermatter{a}.

 The surface integral is
most conveniently evaluated in singular gauge, in which\foot{This is not
true in regular gauge, where $\uAcl$ has nontrivial spatial structure
even at infinity.}
$\uAcl\rightarrow\vhiggs^a\tau^a/2$ so that
 $\uAbar$ can  simply be replaced by $\vbarhiggs^a\tau^a/2$. Singular
gauge is defined by
\eqna\sing
$$\eqalignno{
\uvcl_m\ &=\ {2\over g}\,{\rho^2\over x^2(x^2+\rho^2)}\,
\etabar^{\prime a}_{ml}\,x_l\,{\tau^a\over2}
&\sing a
\cr
\uvcl_{mn}\ &=\ {4\rho^2\over g}\,{1\over x^2(x^2+\rho^2)^2}\,
\big(\,-x^2\etabar^{\prime a}_{mn}+2x_lx_n\etabar^{\prime a}_{ml}
+2x_lx_m\etabar^{\prime a}_{ln}\,\big)\cdot{\tau^a\over2}\qquad&\sing b
\cr
\uAcl\ &=\ {x^2\over x^2+\rho^2}\,\vhiggs^a\,{\tau^a\over2}&\sing c
\cr
\D_l\,\uAcl\ &=\ {2\rho^2\over(x^2+\rho^2)^2}\,x_m\,\vhiggs^b\,
\etabar^{\prime b}_{ln}
\etabar^{\prime a}_{mn}\,{\tau^a\over2}\ .&\sing d
}$$
The gradient of the Higgs field \losuper{b}
including both the classical and fermion-bilinear contributions is
easily obtained, and gives the well-known result\foot{See
Refs.~\refs{\NSVZone-\Amati}. Such expressions for the action are
 of course the source of the familiar lore that the integration
 over instanton scale size is typically dominated by $\rho^2$ on the
order of $1/|\vhiggs|^2,$ at least for $m$-point functions with
$m\ll1/g^2$. Surprisingly, for certain types of supersymmetric correlators,
although not the ones considered in this paper,
it is actually the \it zero\rm-size instantons which control the 
physics \NSVZone.}
\eqn\surfaceaction{\eqalign{S_\higgs+S_\yuk\ &=\ 4\pi^2|\vhiggs|^2\rho^2
\Big(\,1+4\sqrtwo\,
g^{-1}\,{\vbarhiggs^a\over|\vhiggs|^2}\,
\etabar^{\prime a}_{mn}\,\etabar'\sigmabar^{mn}\etabar\,\Big)\cr
 &=\ 
4\pi^2|\vhiggs|^2\rho^2(1-4i\Thetabar\etabar)\ .}}

\subsec{Next-leading-order calculation}

Returning once again to Eq.~\losuper{a}, we note
 that the entire right-handed gauge
superfield $\uWbar_{\dot\alpha}=\big(\ulambdabar\,,\,\uD\,,$ $\sigmabar^{mn}
\uv_{mn}\,,\,\Dbarslash\ulambda\,\big)$ is identically zero;
the $N=1$ gauge components of the
superinstanton live exclusively in the left-handed superfield
$\uW^\alpha$ at leading order in $g^2\rho^2|\vhiggs|^2.$ 
We now refine the superinstanton to next-leading order in
$g^2\rho^2|\vhiggs|^2,$ and verify that $\uWbar_\dalpha$
 turns on at this order. The first step is to improve the choice of
reference configuration $\Psi^\zero.$ As discussed above, the strategy
at this order is to freeze the $N=1$ matter components at their earlier
values, but choose gauge components so that the Euler-Lagrange equations
\eulergauge{}\ hold true. The improved initial choices are thus dictated
by the equations
\eqn\lambdabarfix{\quad\Dslash\ulambdabarzero\ =\ 
\sqrtwo g\,[\,\uAbarcl\,,\,\upsi^\zero\,]\ \equiv\
\sqrtwo g\,[\,\uAbarcl\,,\,\xi'\sigma^{mn}\uv^\cl_{mn}\,]}
and 
\eqn\vmnfix{\D^m\uv^\zero_{mn}\ 
 =\ 
-2ig\,[\,\uAbarcl\,,\D_n\uAcl\,]}
with $\ulambda^\zero$ and $\uD^\zero$ still zero. Repeating the earlier
two-step sweeping-out procedure
then gives the improved gauge components of the superinstanton,
which supersede Eq.~\losuper{a}:
\eqna\improving
$$\eqalignno{
\uv_{mn}\ &=\ \uv_{mn}^\zero+i\xi(x)\sigma_{[\,n}\D_{m\,]}
\ulambdabar^\zero+4i\ulambdabar^\zero
\sigmabar_{mn}\etabar
\cr&\qquad
+2ig\,\xi(x)^2\,\ulambdabar^\zero
\sigmabar_{mn}\ulambdabar^\zero\ ,&\improving a
\cr
\ulambda\ &=\ -\xi(x)\sigma^{mn}\uv_{mn}^{\SDzero}+i\xi(x)^2\,
\Dslash\ulambdabar^\zero\ ,&\improving b
\cr
\ulambdabar\ &=\ \ulambdabarzero \ +\
\sigmabar^{mn}\Thetabar\uv_{mn}^{\ASDzero}\ ,&\improving c
\cr
\uD\ &=\  -\xi(x)\Dslash\big( \ulambdabarzero \ +\
\sigmabar^{mn}\Thetabar\uv_{mn}^{\ASDzero}\big)\ .&\improving d}$$

We need to verify that the expression \improving{c} for $\ulambdabar$ 
is actually related by $SU(2)_\R$ symmetry \SUtwoRb\ to 
the expression \losuper{c} for $\upsibar$, despite appearances.
This is easily accomplished, first by decomposing the field strength
into self-dual and anti-self-dual components, $\uv_{mn}^\zero=
\uv^\SDzero_{mn}+\uv^\ASDzero_{mn},$ and next by recasting the
Bianchi identity (\moreuseful$d$) as
$\D^m\uv^\SDzero_{mn} = \D^m\uv^\ASDzero_{mn}.$
This fact, together with Eqs.~\sconfrewrite\ and (\useful $c$),
then allows us to rewrite Eq.~\vmnfix\ as
\eqn\vasdfinal{\eqalign{\Dslash\big(\sigmabar^{mn}\Thetabar
\uv_{mn}^\ASDzero\big)\ &=\ 2ig\,[\,\uAbarcl\,,\,-\Thetabar
\Dbarslash\uAcl\,]\cr&=\ \sqrtwo g\,[\,\uAbarcl\,,\,
-(x^{}_k-x^0_k)\etabar'
\sigmabar^k\sigma^{mn}\uv_{mn}^\cl\,]\ .}}
Adding together Eqs.~\lambdabarfix\ and \vasdfinal\ yields the following
condition for
 the right-hand side of Eq.~\improving{c}:
\eqn\lambdabarrhs{\Dslash\ulambdabar\ \equiv\ \Dslash\big(\ulambdabarzero  +
\sigmabar^{mn}\Thetabar\uv_{mn}^{\ASDzero}\big)\ =\ 
\sqrtwo g\,[\,\uAbarcl\,,\,\xi'(x)\sigma^{mn}\uv_{mn}^\cl\,]\ ,}
where $\xi'(x)$ was defined in Eq.~\etabarprimedef{a}. Save
for the switch $\xi(x)\rightarrow\xi'(x)$ this is \it precisely \rm
Eq.~\eulerantimatter{b} for $\upsibar,$ from which we conclude
\eqn\lambdabaranalog{\qquad\ulambdabar\ =\ -i\sqrtwo
\big(\xi'(x)\Dslash+2\etabar'\big)\uAbarcl 
\ ,\quad
\ulambdabar^\zero\ =\ \ulambdabar\,{\Big|}_{\etabar'=0}\ =\ 
-i\sqrtwo\xi'\Dslash\uAbarcl }
by analogy with Eq.~\losuper{c}.
$SU(2)_\R$ symmetry for the antifermions is now manifest, as is the validity
of the Euler-Lagrange equations \eulergauge{c,d}; the reader can verify
that Eqs.~\eulergauge{a,b} are true as well, up to still higher-order
perturbative corrections.

To complete the determination of the short-distance
superinstanton through this order,
it remains only to specify $\uv_{mn}^\zero.$ The anti-self-dual part
is given by $\sigmabar^{mn}\Thetabar\uv_{mn}^{\ASDzero} = 
 \ulambdabar\,{\big|}_{\xi'=0}$. The self-dual piece has the general
form $\uv_{mn}^{\SDzero}=\uv_{mn}^\cl+\delta\uv_{mn}^{\SDzero}$; here
$\delta\uv_{mn}^{\SDzero}$ is a relative correction of order $g^2\rho^2
|\vhiggs|^2$ to the BPST field strength, but as it is only determined
up to an admixture of bosonic zero modes, explicit
expressions are not particularly illuminating, nor are they needed
in what follows.

\newsec{\bf The Seiberg-Witten Effective Action at the 1-Instanton Level}

We now apply these results to some explicit calculational checks of the
Seiberg-Witten effective Lagrangian, Eqs.~\FSWdef\ and
\Leffcomp. We will evaluate
the 1-instanton contribution to the family of $(l+4)$-point 
functions\foot{As is obvious from Eqs.~\FSWdef\ and \Leffcomp, these chirality
violating 4-fermi Green's  functions
receive their leading contribution, not from
${\cal F}_{\rm inst}'''',$ but from ${\cal F}_{\rm pert}''''.$
Diagrammatically, this perturbative contribution sums the
one-loop polygon graphs with four massless fermions plus $l$ 
 fluctuating anti-Higgses on the
external legs, and massive quanta running around the polygon. Our point of
view is that understanding the instanton series is important theoretically,
even
 if in weak coupling the instanton contribution to individual cross sections
is negligible. The reason that one examines instanton correlators
of antifermions rather than fermions is simply that,
 while the latter  appear at one lower
order in perturbation theory in $g$, they fall off one power
more slowly with $x$. Thus in the long-distance domain $x\gg1/M_W$ the fermion
Green's functions
are actually sub-dominant, and contribute to the higher-derivative
corrections to Eq.~\Leffdef\ 
(vice versa in \it anti\rm-instanton backgrounds) \ADSone. }
\eqn\familypoint{\langle\,
\lambdabar_{\dot\alpha}(x_1)\,
\lambdabar_{\dot\beta}(x_2)\,
\psibar_{\dot\gamma}(x_3)\,
\psibar_{\dot\delta}(x_4)\,
\delta\Abar(x_5)\cdots\delta\Abar(x_{l+4})\,\rangle\ ,}
extending slightly the analysis of Finnell and Pouliot \FPone\ (for
higher gauge groups, see Ref.~\ISone). However,
we will also evaluate the $(l+3)$-point and $(l+2)$-point functions
\eqn\secondfamily{\langle\,v_{mn}(x_1)\,
\lambdabar_{\dot\alpha}(x_2)\,
\psibar_{\dot\beta}(x_3)\,
\delta\Abar(x_4)\cdots\delta\Abar(x_{l+3})\,\rangle\ }
and
\eqn\thirdfamily{\langle\,v_{mn}(x_1)\,v_{kl}(x_2)
\delta\Abar(x_3)\cdots\delta\Abar(x_{l+2})\,\rangle\ ,}
both of which require the extra machinery developed above. 

The first task is to extrapolate the relevant long-distance effective
$U(1)$ fields from the short-distance singular-gauge superinstanton,
which is summarized in Eqs.~\losuper{b,c}$\,$, \sing{}, \improving{},
and \lambdabaranalog. The simplest field  is the anti-Higgs,
which, unlike the Higgs, is free of fermion bilinears
through subleading order:
\eqn\Abarexpand{\uAbar\, =\, \uAbarcl\, =\, {(x-x_0)^2\over
(x-x_0)^2+\rho^2}\,\vbarhiggs^a\,{\tau^a\over2}
\, =\,
\Big(\,1-{\rho^2\over(x-x_0)^2}+{\rho^4\over(x-x_0)^4}-\cdots\Big)
\,\vbarhiggs^a\,{\tau^a\over2}\ .}
This Taylor expansion is strictly valid only in the regime
$\rho^2\ll(x-x_0)^2\ll1/M_W^2.$ Notice, however,
 that the first two terms (but not the higher terms) also
satisfy the long-distance
Euler-Lagrange equations \longeom, which are valid for $(x-x_0)^2\gg1/M_W^2$.
  This illustrates the ``patching
condition'' discussed in Sec.~3. Accordingly, up to $O(g^2)$ corrections,
 we equate the long-distance tail of  $\Abar$ with the truncated
expression $\vbarhiggs\,-\,\vbarhiggs
\rho^2/(x-x_0)^2\,.$ In other words (subtracting off the vev),
\eqn\delAbarlong{\delta\Abar(x)\ =\ -{\vbarhiggs\rho^2\over(x-x_0)^2}\ 
=\ -\vhiggs^{-1}\,S_\higgs\cdot G(x,x_0)\ ,}
where $G(x,x_0)=1/4\pi^2(x-x_0)^2$ is the massless Euclidean propagator. 

So now  consider each of the three families of Green's functions 
in turn, starting with \familypoint. 
The  1-instanton measure \tHooft\ in the $N=2$ \susic\ gauge
theory has the well-known form:
\eqn\measure{\int 2^{10}\pi^6g^{-8}(\mu\rho)^8d^4x_0\,{d\rho
\over\rho^5}\times\Big({16\pi^2\mu\over g^2}\Big)^{-2}d^2\xi
d^2\xi'\times
\Big({32\pi^2\rho^2\mu\over g^2}\Big)^{-2}d^2\etabar d^2\etabar'\times
e^{-S_\inst}\ .}
The factors in big parentheses contain the norms of the \susic\ and
superconformal fermion zero modes, respectively (see Appendix A).
Since $S_\inst$ does not depend on 
$\xi$ and $\xi'$, these Grassmann integrations must be
saturated by the antifermions in \familypoint. In terms of the
spinor propagator $S(x,x_0)\equiv\delslash G(x,x_0)$, the $\xi$ and $\xi'$
components of the antifermions
are\foot{Recall that the long-distance fields are neutral
under the unbroken $U(1)$ so that $\D_n\rightarrow\partial_n$.
Here and in Eq.~\delAbarlong, we choose to express the right-hand
sides in terms of $S_\higgs,$ in order that these expressions may be
immediately promotable to the multi-instanton case, where $S_\higgs$
is much more complicated.}
\eqn\psibarlong{\psibar_\dalpha(x)\ =\ -i\sqrtwo\xi^\alpha\delslash_{\alpha
\dalpha}\Abar_\cl(x)\ =\ i\sqrtwo\vhiggs^{-1}S_\higgs\,
\xi^\alpha S_{\alpha\dalpha}(x,x_0)}
and likewise
\eqn\lambdabarlong{\lambdabar_\dalpha(x)\
 =\ i\sqrtwo\vhiggs^{-1}S_\higgs\,\xi^{\prime\alpha} S_{\alpha\dalpha}
(x,x_0)}
as follows from Eqs.~\losuper{c}, \lambdabaranalog, and \delAbarlong.
These expressions, too, satisfy the long-distance equations
\longeom\ and are therefore dictated by the ``patching condition.'' 
Integrating out the (lifted) superconformal modes from the
measure gives
\eqn\sconfint{\int d^2\etabar d^2\etabar'\,\exp(-S_\yuk)\ =\
-2^9\pi^4g^{-2}\rho^4\vbarhiggs^2\ .}
The remaining integrations are elementary, and yield 
\eqn\newelement{\eqalign{\int d^4x_0&
\epsilon^{\alpha\beta}\,S_{\alpha\dalpha}(x_1,x_0)S_{\beta\dbeta}(x_2,x_0)\,
\epsilon^{\gamma\delta}\,S_{\gamma\dgamma}(x_3,x_0)S_{\delta\ddelta}(x_4,x_0)
\cr&\times\ G(x_5,x_0)\times\cdots\times G(x_{l+4},x_0)\ 
{(5+l)!\,\Lambda^4\over16\pi^2g^2(-\vhiggs)^{6+l}}\ ,}}
with $\Lambda$ as in Eq.~\FPdef.
This position-space Green's function may be interpreted as coming  from 
an instanton-induced effective local vertex
\eqn\effvertex{\sum_{l=0}^\infty\ {1\over2!\,2!\,l!}\,\psi^2\lambda^2
(\delta A)^l\
{(5+l)!\,\Lambda^4\over16\pi^2g^2(-\vhiggs)^{6+l}}\ =\
{15\Lambda^4\over8\pi^2g^2}\,
{\psi^2\lambda^2\over A^6}\ .}
The binomial theorem has been used to reconstitute in the denominator
the total Higgs field $A(x_0)=\vhiggs+\delta A(x_0)$.
It is interesting to see how the combinatorics of the collective
coordinate integration promotes $\vhiggs^{-6}$ to $A^{-6}$, as
required by the Seiberg-Witten effective action.

Next we look at the  Green's functions \secondfamily\ and \thirdfamily. In
order to saturate the $\xi$ and $\xi'$ integrations we need to extract
the piece of $v_{mn}$ bilinear in $\xi$ and $\xi'.$ Here
our careful sweeping-out procedure in Sec.~4.4 pays off: the second term
on the right-hand side of Eq.~\improving{a}
contains just such a piece, which we write as
\eqn\bilinearvmn{\sqrtwo\vhiggs^{-1}S_\higgs\,\xi
\sigma_{[\,n}\delbarslash\,\partial_{m\,]}\,G(x,x_0)\xi'\ ,}
using Eqs.~\lambdabaranalog\ and \lambdabarlong. This expression, too,
 satisfies the long-distance equations \longeom. It is also purely
anti-self-dual (as pointed out to us by Yung), meaning that it
fits into the right-handed
superfield ${\overline W}_\dalpha$, rather than in $W^\alpha$ where
 $v_{mn}^\cl$ itself lives.

 We can reexpress Eq.~\bilinearvmn\ in a more illuminating way, as
\eqn\newbi{-\sqrtwo\vhiggs^{-1}S_\higgs\,\xi\sigma^{kl}\xi'\,
G_{mn,kl}(x,x_0)\ ,}
where $G_{mn,kl}$ is the gauge-invariant propagator of $U(1)$ field strengths:
\eqn\Gmnkldef{G_{mn,kl}(x,x_0)\ =\ \big(
\,\eta_{nl}\partial_m\partial_k-\eta_{nk}\partial_m\partial_l
-\eta_{ml}\partial_n\partial_k+\eta_{mk}\partial_n\partial_l\,\big)G(x,x_0)\
 .}
The families of 
Green's functions \secondfamily\ and \thirdfamily\ then work out to,
respectively,
\eqn\newerelement{\eqalign{\int d^4x_0\,&
\sigma^{kl\,\alpha\beta}G_{mn,kl}(x_1,x_0)\,
S_{\alpha\dalpha}(x_2,x_0)S_{\beta\dbeta}(x_3,x_0)
\cr&\times\ G(x_4,x_0)\times\cdots\times G(x_{l+3},x_0)\ 
{(4+l)!\,\Lambda^4\over16\sqrtwo\pi^2g^2(-\vhiggs)^{5+l}}\ }}
and
\eqn\newestelement{\eqalign{\int d^4x_0\,&
\trtwo\sigma^{pq}\sigma^{rs}\cdot G_{mn,pq}(x_1,x_0)\,
G_{kl,rs}(x_2,x_0)\,
\cr&\times\ G(x_3,x_0)\times\cdots\times G(x_{l+2},x_0)\ 
{(3+l)!\,\Lambda^4\over32\pi^2g^2(-\vhiggs)^{4+l}}\ .}}
These, in turn, correspond to the effective pointlike vertices
\eqn\neweffvertex{\sum_{l=0}^\infty\ {1\over l!}\,\lambda\sigma^{kl}
\psi\,v_{kl}\,(\delta A)^l\
{(4+l)!\,\Lambda^4\over16\sqrtwo\pi^2g^2(-\vhiggs)^{5+l}}\ =\
-{3\Lambda^4\over2\sqrtwo\pi^2g^2}\,
{\lambda\sigma^{kl}\psi v_{kl}\over A^5}\ }
and
\eqn\newesteffvertex{\eqalign{\sum_{l=0}^\infty\ {1\over 2!\,l!}\,
\trtwo\sigma^{pq}\sigma^{rs}\cdot v_{pq}v_{rs}\,(\delta A)^l\
{(3+l)!\,\Lambda^4\over32\pi^2g^2(-\vhiggs)^{4+l}}\ &=\
{3\Lambda^4\over32\pi^2g^2}\,
{\trtwo\sigma^{pq}\sigma^{rs}\cdot v_{pq}v_{rs}\over A^4}\cr& =\
-{3\Lambda^4\over16\pi^2g^2}\,
{ \big(v^\SD_{pq}\big)^2\over A^4}\ ,}}
again with the help of the binomial theorem.

Comparing the three effective vertices \effvertex, \neweffvertex\
and \newesteffvertex\ to their counterparts in the Seiberg-Witten
Lagrangian, Eqs.~\FSWdef\
and \Leffcomp, we find agreement if and only if ${\cal F}_1
=1/2\,,$ confirming their prediction. The remaining vertices in
$\L_\eff,$ namely the renormalized  fermion and Higgs kinetic energies,
share the property that (up to an integration by parts) they vanish
identically when the leading-order equations of motion are used.
This means that they only affect on-shell processes at a higher order
in perturbation theory. In the instanton language, they presumably map onto
propagator corrections in instanton backgrounds, which lie beyond
the scope of this paper.

We now apply the lessons learned at the 1-instanton level to the
study of multi-instantons.

\newsec{ABC's of ADHM}
\subsec{Three Ans$\ddot{a}$tze}

In this Section we give a physicist's
introduction to the ADHM construction of
multi-instantons \refs{\ADHMone-\ADHMtwo}, 
which in our idiosyncratic formulation will be seen to
 rely on three remarkably
simple ans$\ddot{\rm a}$tze. (For additional information, see especially
Refs.~\refs{\Oone,\CWSone,\CGTone}.)
We will specialize to the gauge group $SU(2)$ from the outset, and we  will
also preserve the distinction between dotted and undotted $SU(2)$
indices in order to facilitate supersymmetrization. In fact (in this
Section only) we shall often exhibit \it all \rm indices (five
different types!\foot{Apart from the $SU(2)$ indices $\alpha, \beta,
\dots$ and $\dalpha,\dbeta,\dots$ and Lorentz indices $m,n$ there
are also Greek and Roman
ADHM matrix indices $\kappa, \lambda = 0,1,\cdots,n$ and
$k,l=1,2,\cdots,n$ where $n$ is the winding number (number of instantons).
Additional conventions for the remainder of the paper:
each Lorentz vector $z_m$ has associated with it \it quaternions \rm
$z_{\alpha\dalpha}=z_m\sigma^m_{\alpha\dalpha}$ and 
$\bar z^{\dalpha\alpha}=z^m\sigmabar_m^{\dalpha\alpha}$. For quantities
with two undotted (likewise two dotted) $SU(2)$ indices, overbarring
reverses the sign of the $\sigma_\alpha{}^\beta$ components but not
of the $\delta_\alpha{}^\beta$ components, so that $\bar z_\alpha{}^\beta
=\delta_\alpha{}^\beta\trtwo z-z_\alpha{}^\beta$; if a quantity
also carries ADHM indices then the overbar transposes in these as well
(overbarring also conjugates any multiplicative complex phase).
In contrast the superscript $\scriptstyle T$ transposes  only 
in the ADHM matrix indices, whereas
the symbol $\trtwo$  traces only  over the dotted or undotted $SU(2)$
indices. 
Note the very useful identity $z_{\alpha\dalpha}\bar y^{\dalpha\beta}
+y_{\alpha\dalpha}\bar z^{\dalpha\beta}=\trtwo z\bar y$; more generally,
for a quantity $X_\alpha{}^\beta$ that carries ADHM indices one has
$\trtwo X=X+\bar X^T.$ We define $|z|^2=z\bar z=\bar zz$ for any quantity
not having ADHM indices, be it $z_{\alpha\dalpha},$ $z_\alpha{}^\beta$
or $z^\dalpha{}_\dbeta.$ Note that $|z_{\alpha\dalpha}|^2=-z_nz^n$ due
to the Wess and Bagger metric. Unit-normalized quaternions,
$|z|^2=1,$ are elements of $SU(2)$. Finally note that 
$z^{\alpha\beta}=z^{\beta\alpha}-\epsilon^{\beta\alpha}\trtwo z$
whereas
$z^{\dalpha\dbeta}=z^{\dbeta\dalpha}+\epsilon^{\dbeta\dalpha}\trtwo z$.
})
for maximal clarity.

An heuristic motivation for the ADHM construction is to notice an
interesting pattern: just
as a zero-instanton solution of pure $SU(2)$ Yang-Mills theory is 
by definition a pure gauge,
\eqn\puregauge{v_n{}^\dalpha{}_\dbeta\ =\ \Ubar^{\dalpha\alpha}\partial_n
U_{\alpha\dbeta}\ ,\qquad \Ubar^{\dalpha\alpha}U_{\alpha\dbeta}\ =\ 
\delta^\dalpha{}_\dbeta\ ,}
so too a 1-instanton configuration is expressible as
\eqn\adhmansatz{v_n{}^\dalpha{}_\dbeta\ =\ 
\Ubar_\lambda^{\dalpha\alpha}\partial_n
U_{\lambda\,\alpha\dbeta}\ ,\qquad 
\Ubar_\lambda^{\dalpha\alpha}U_{\lambda\,\alpha\dbeta}\ =\ 
\delta^\dalpha{}_\dbeta\ ,}
where the new index $\lambda$ is summed over 0 and 1. For example, the
usual singular-gauge instanton \sing{a} follows from the choice
\eqn\adhmsing{U_{0\,\alpha\dbeta}\ =\ 
\sqrt{x^2\over x^2+\rho^2}\cdot\sigma_{0\alpha\dbeta}\ ,\qquad 
U_{1\,\alpha\dbeta}\ =\ 
-{\rho\over x^2}
\sqrt{x^2\over x^2+\rho^2}\cdot x_{\alpha\dalpha}\bar u^{\dalpha\beta}
\sigma_{0\beta\dbeta}}
while the regular-gauge instanton follows from
\eqn\adhmreg{U_{0\,\alpha\dbeta}\ =\ 
-{1\over\sqrt{x^2+\rho^2}}\cdot u_{\alpha\dalpha}\bar x^{\dalpha\beta}
\sigma_{0\beta\dbeta}\ 
,\qquad 
U_{1\,\alpha\dbeta}\ =\
\sqrt{\rho^2\over x^2+\rho^2}\cdot\sigma_{0\alpha\dbeta}\ ,}
where in quaternionic notation $u_{\alpha\dalpha}\equiv
u_n\sigma^n_{\alpha\dalpha}$ is an iso-orientation matrix in
the spin$\hbox{-}1/2$ representation of $SU(2)$, i.e., satisfying
$|u|^2=1.$ 
ADHM's first key ansatz for where to search for the $n$-instanton solutions
is, naturally enough, Eq.~\adhmansatz\ again, where now 
$\lambda=0,1,\cdots,n$.

To find a solution of the Yang-Mills equations it suffices to establish
self-duality of the field strength\foot{The shortest imaginable proof
 that (anti)self-dual gauge fields automatically
satisfy the Yang-Mills equations of motion uses the Bianchi identity
(\moreuseful$d$):
$\D_m\uv^{mn}\ \propto\ \D_m\epsilon^{mnpq}\uv_{pq}\ =\ 0$.
Henceforth, rather than $\uv_n=v_n^a\tau^a/2$ as before, the $SU(2)$ gauge
field will be denoted $v_n{}^\dalpha{}_\dbeta$, and obeys
$\trtwo v_n{}^\dalpha{}_\dbeta\equiv v_n{}^\dalpha{}_\dalpha=0$; likewise
for the other fields. Furthermore, 
following the ADHM tradition we will work with anti-Hermitian gauge fields,
and set $g=1$ in the following.}
\eqn\adhmvmn{v_{mn}{}^\dalpha{}_\dbeta\ \equiv\ 
\partial_{\,[\,m}v_{n\,]}{}^\dalpha{}_\dbeta\ +\ 
v_{[\,m}{}^\dalpha{}_\dgamma\,
v_{n\,]}{}^\dgamma{}_\dbeta\ =\ \partial_{\,[\,m}\Ubar^{\dalpha\alpha}_\lambda
\,\big(\,\delta_{\lambda\kappa}\delta_\alpha{}^\beta-
\P_{\lambda\kappa\,\alpha}{}^\beta\,\big)\,
\partial_{n\,]}U_{\kappa\,\beta\dbeta}}
where 
\eqn\projdef{\P_{\lambda\kappa\,\alpha}{}^\beta\ =\ 
 U_{\lambda\alpha\dalpha}\Ubar_\kappa^{\dalpha\beta}\ .}
Note that $\P$, and hence also $1-\P,$ is a projection operator, satisfying
\eqn\projconds{0\ =\ (1-\P) U\ 
\ =\ \Ubar(1-\P)\ \ ,\qquad \P\,=\,\bar\P\ ,\qquad\P^2=\P\ .}
ADHM's second key ansatz is to assume that $1-\P$ can be factorized as 
\eqn\Deltadef{\delta_{\lambda\kappa}\delta_\alpha{}^\beta-
\P_{\lambda\kappa\,\alpha}{}^\beta \ =\
\Delta_{\lambda l\,\alpha\dalpha}\,f_{lk}{}^\dalpha{}_\dbeta\,
\Deltabar^{\dbeta\beta}_{k\kappa}}
for some matrices $\Delta$ and $f$, and that Eq.~\projconds\ is satisfied
by virtue of the more basic conditions
\eqn\perpspace{0\ =\ \Deltabar^{\dbeta\beta}_{k\kappa}\,U_{\kappa\beta\dalpha}
\ =\ \Ubar^{\dbeta\alpha}_\lambda\,\Delta_{\lambda l\,\alpha\dalpha}\ ,
\qquad f=\fbar\ .}
In a moment we will solve for $\Delta_{\lambda l},$ but already we know
its size: by comparing the dimensions of the nullspaces of $U\bar U$ and
$1-\Delta f\bar\Delta$, we conclude that $\Delta_{\lambda l}$ must be
a rectangular quaternion-valued matrix of dimension $(n+1)\times n.$ 
We will refer to both $\lambda=0,\cdots,n$ and $l=1,\cdots,n$ as
``ADHM indices.''

Returning to the issue of self-duality, we can now rewrite 
Eq.~\adhmvmn\ as
\eqn\adhmvmnb{v_{mn}{}^\dalpha{}_\dbeta\ =\ 
\Ubar^{\dalpha\alpha}_\lambda \,
\partial_{\,[\,m}
\Delta_{\lambda l\,\alpha\dgamma}\,f_{lk}{}^\dgamma{}_\ddelta \,
\partial_{n\,]}
\Deltabar^{\ddelta\beta}_{k\kappa}\,
U_{\kappa\,\beta\dbeta}\ ,}
 with  an integration by parts.
ADHM's third and final key ansatz is now easily anticipated: If
one can arrange that 
$\partial_m
\Delta_{\lambda l\,\alpha\dalpha}=\hbox{stuff}\times\sigma_{m\beta\dalpha}$
and also that  $\sigma_m$ commutes through $f$, then
the right-hand side of Eq.~\adhmvmnb\ will be of the form 
$\hbox{stuff}\times\sigma_{mn}\times\hbox{stuff}$, 
and self-duality is guaranteed (being a built-in property of
$\sigma_{mn}$). So we postulate, first,
 that $\Delta$ is a linear left-multiplied
function of $x_{\beta\dalpha}\equiv x^m\sigma_{m\beta\dalpha},$
\eqn\Deltapostulate{\Delta_{\lambda l\,\alpha\dalpha}
\ =\ a_{\lambda l\,\alpha\dalpha}\ +\
b_{\lambda l\,\alpha}{}^\beta\,x_{\beta\dalpha}}
where $a_{\lambda l}$ and $b_{\lambda l}$ are $(n+1)\times n$-dimensional
matrices of constant quaternions; and second
\eqn\fidentity{f_{lk}{}^\dalpha{}_\dbeta\ =\ f_{lk}\,\delta^\dalpha{}_\dbeta}
so that it commutes with the Pauli matrices. We then find, as promised,
\eqn\vmnagain{v_{mn}{}^\dalpha{}_\dbeta\ =\ \big(
v_{mn}{}^\dalpha{}_\dbeta\big)^{\rm\sst dual}\ =\ 
4\Ubar_\lambda^{\dalpha\alpha}
\,b_{\lambda l\,\alpha}{}^\beta\,\sigma_{mn\,\beta}{}^\gamma
\,f_{lk}\,\bbar_{k\kappa\,\gamma}{}^\delta\,
U_{\kappa\,\delta\dbeta}\ .}
That the winding number is in fact $n$ is checked in Sec.~7.4 below.

\subsec{Solving the ADHM equations}

At this point no further ans$\ddot{\rm a}$tze are needed;
 the problem of solving the Yang-Mills equations in the $n$-instanton
sector has been reduced to constructing ADHM matrices
$f_{lk},$ $U_{\lambda\alpha\dalpha}$, $b_{\lambda l\,\alpha}{}^\beta,$
and $a_{\lambda l\,\alpha\dalpha}$  such that \projdef-\perpspace\ 
are a consistent set of equations. Let us solve for each of these 
four quantities in turn.
First, these equations imply that $\P\Delta=0$;
therefore  $f$ can be eliminated in terms of $\Delta$, via
\eqn\Deltaid{\Deltabar^{\dbeta\beta}_{k\kappa}\,\Delta_{\kappa l\,
\beta\dalpha}\ =\ \big(\finv\big)_{kl}\,\delta^\dbeta{}_\dalpha\ .}
Of course this condition only makes sense if the left-hand side really
is proportional to $\delta^\dbeta{}_\dalpha$, which is tantamount
to requiring
\eqna\crucial
$$\eqalignno{\abar a\ &=\ (\abar a)^T\ \propto\ \delta^\dbeta{}_\dalpha\
,&\crucial a
\cr
\bbar a\ &=\ (\bbar a)^T\ ,&\crucial b\cr
\bbar b\ &=\ (\bbar b)^T\  \propto\ \delta_\alpha{}^\beta\
,&\crucial c}$$
as follows from a Taylor expansion in $x$ (the superscript
$\scriptstyle T$ denotes a transpose in the ADHM indices only).
We will return to these conditions shortly.

Second, let us solve for $U_\lambda,$ and hence $v_n$ itself,
in terms of $\Delta.$
Equating the two alternative expressions \projdef\ and \Deltadef\ for $\P$
and setting $\kappa=0$ implies, for $\lambda=0$ and $\lambda\neq0$
respectively:
\eqn\Uzeroeom{|U_0|^2\ =\ 1-\hf f_{lk}\trtwo w_l\wbar_k\ ,\qquad
w_{l\alpha\dalpha}=\Delta_{0l\,\alpha\dalpha}}
and
\eqn\Ulambdaeom{U_\lambda\ =\ -{1\over|U_0|^2}\,\Delta_{\lambda l}\,
f_{lk}\,\wbar_k\,U_0\ ,\qquad\lambda\neq0\ .}
Crucially,
thanks to \Deltaid, these solutions are consistent: the expressions \projdef\
and \Deltadef\ automatically remain equal even when both $\kappa$
and $\lambda$ are nonzero. Note that there are an
 infinite number of quaternions $U_0$ satisfying
Eq.~\Uzeroeom, but these
 are all equivalent up to gauge transformations of $v_n$.
Generalizing the singular-gauge expression \adhmsing\ which served us well
 in the 1-instanton sector, we will specify 
\eqn\Uzerogauge{U_{0\alpha\dalpha}\ =\ \sigma_{0\alpha\dalpha}\,
\big(1-\hf f_{lk}\trtwo w_l\wbar_k\big)^{1/2}}
in what follows, whenever an explicit choice is called for.

Third, let us eliminate the degrees of freedom of 
$b_{\lambda l\,\alpha}{}^\beta$ entirely from the
problem. 
To this end, it is helpful to catalog the complete set of invariances
of the ADHM construction. The usual $SU(2)$ gauge transformations
of $v_n$ read
\eqn\inserta{U_{\lambda\alpha\dalpha}\rightarrow
U_{\lambda\alpha\dbeta}\,\Omega(x)^\dbeta{}_\dalpha\ ,\qquad
\Delta\rightarrow\Delta\ ,\qquad f\rightarrow f}
where $\bar\Omega^\dalpha{}_\dgamma\Omega^\dgamma{}_\dbeta=\delta^\dalpha
{}_\dbeta\,$. For example, the gauge transformation between \adhmsing\
and \adhmreg\ is induced by
\eqn\insertb{\Omega(x)^\dbeta{}_\dalpha\ =\ 
-{1\over\sqrt{x^2}}\,\sigmabar_0^{\dbeta\beta}u_{\beta\dgamma}\,\bar
x^{\dgamma\gamma}\sigma_{0\gamma\dalpha}\ .}
In constrast, consider the two sets of transformations
\eqn\insertc{\Delta_{\lambda l\beta\dbeta}\rightarrow
\Delta_{\lambda k\beta\dbeta}\,B_{kl}\ ,\qquad
f\rightarrow B^{-1}\cdot f\cdot(B^{-1})^T\ ,\qquad
U\rightarrow U}
and 
\eqn\insertd{\Delta_{\lambda l\beta\dbeta}\rightarrow
\Lambda_{\lambda\kappa\,\beta}{}^\alpha
\Delta_{\kappa l\alpha\dbeta}\ ,\qquad
f\rightarrow f\ ,\qquad
U_{\lambda\beta\dbeta}\rightarrow
\Lambda_{\lambda\kappa\,\beta}{}^\alpha
U_{\kappa\alpha\dbeta}\ ,\qquad\bar\Lambda\Lambda=1\ ,}
where $B$ and $\Lambda$ are independent of $x$. While
these obviously preserve the
various ADHM relations and constraints detailed above, 
they have no effect on the
gauge field \adhmansatz\ itself, hence they commute with
gauge fixing. We now exploit these two invariances to simplify $b$, as
follows. With an initial $\Lambda$ whose top row lives in the $\perp$-space
of $b$, $b$ can be brought into the form
$$b\ =\ \pmatrix{0&\cdots&0\cr{}&{}&{}\cr
{}&b'&{}\cr{}&{}&{}\cr}\ .$$
Since the product 
$\bbar'b'$ is a real symmetric \it scalar\rm-valued $n\times n$
matrix as follows from Eq.~\crucial{c}, it can be factored as
$O\cdot\mu\cdot O^T$, where $O$ is an orthogonal matrix, and ${\mu}$ is 
a diagonal matrix of nonnegative eigenvalues. So next we let $\Delta\rightarrow
\Delta B$ as per \insertc, with $B=O\cdot{\mu}^{-1/2}$. This
ensures that the new $b'$ is unit-normalized:
$\bbar'_{kk'\alpha}{}^\beta\, b'_{k'l\,\beta}{}^\gamma=\delta_{kl}
\delta_\alpha{}^\gamma\,.$ Applying a follow-up unitary transformation
\insertd\ with
$$\Lambda\ =\ \pmatrix{1&0&\cdots&0\cr
0&{}&{}&{}\cr
\vdots&{}&\bbar'&{}
\cr0&{}&{}&{}\cr}\ ,$$
we then rotate $b$ into its final
canonical form, while fixing our notation for the constant matrix
$a$ as follows:
\eqn\bcanonical{b_{\lambda k}\ =\ 
\pmatrix{0&\cdots&0 \cr \delta_\alpha{}^\beta & \cdots & 0 \cr
\vdots & \ddots & \vdots \cr 0 & \cdots &
\delta_\alpha{}^\beta}
\quad,\quad
a_{\lambda k}\ =\ \pmatrix{w_1&\cdots&w_n\cr{}&{}&{}\cr
{}&a'_{lk}&{}\cr{}&{}&{}}\ .}
Here $a'_{lk\,\alpha\dalpha}$ is an $n\times n$ matrix of quaternions.

Fourth, we discuss the remaining unknown,  $a_{\lambda k}$ itself, which is
now constrained \it solely \rm by the two conditions
\crucial{a,b}. Notice that these constraints  are invariant under the linear
shifts 
\eqn\globaltrans{a\ \rightarrow\ a\,-\,bx_0\ .}
So the degree of freedom of $a$ that is proportional to $b$ itself
has a special role to play: that of the \it position \rm of the
multi-instanton (in our notation, this is $-x_0$). 
With the canonical choice for $b$, Eq.~\crucial{b}  simply means
\eqn\newacond{a'\ =\ a^{\prime\, T}\ .}
Equation~\crucial{a} is more complicated; it 
defines a set of coupled quadratic
constraints on the quaternionic elements of $a$.
For $n=1$ it is automatically satisfied; for $n=2$ it is  easily solved
\CWSone\ (as reviewed
 in Sec.~8.1 below); and for $n=3$ an intricate solution has
been constructed in
Ref.~\CWSone. However for $n>3$ no solution of this constraint is known.
This is arguably the 
biggest failing of the ADHM construction, and has hampered progress
in the study of multi-instantons, particularly as regards the construction
of the moduli space integration measure.

\def\vW{\vec w}
Let us pause to count the independent parameters of the quaternion-valued
matrix $a_{\lambda k}.$ Thanks to \newacond\ it contains $2n(n+3)$
scalar degrees of freedom. Irrespective of the existence of an explicit
solution, Eq.~\crucial{a} then imposes $\textstyle{3\over2}
n(n-1)$ constraints on the upper-triangular traceless quaternionic elements of
$\abar a,$ leaving $\hf n(n+15)$ degrees of freedom. But for $n>1$ this
is still  too large a set; by considering the limit of $n$  widely separated
distinguishable instantons we see that the correct number should
of course be $8n$ ($4n$ positions, $n$ scale sizes, and $3n$ iso-orientations).
In other words, we are still lacking $\hf n(n+15)-8n=
\hf n(n-1)$ constraints.
To see where these come from, look again at Eqs.~\insertc-\bcanonical.
Even after one fixes the canonical form \bcanonical\ for $b$
and chooses a particular $SU(2)$ orientation for $U_0$ (e.g.,
Eq.~\Uzerogauge), there are
still   $x$-independent $O(n)$ transformations
which act nontrivially on $a,$ namely:
\eqn\trans{\Delta\,\rightarrow\
\pmatrix{1&0&\cdots&0\cr0&{}&{}&{}\cr\vdots&{}&R^T&{}\cr0&{}&{}&{}\cr}
\cdot\Delta\cdot R\ ,\quad f\rightarrow R^T\cdot f\cdot R\ ,
\quad U_0\rightarrow U_0\ ,\quad U_l\rightarrow R_{kl}U_k}
where $R\in O(n)$ and carries no $SU(2)$ indices. Obviously this is a composite
of transformations of the type \insertc\ and \insertd, specifically
arranged to preserve the canonical form of $b$ and $U_0$. 
As such, it too commutes with ordinary gauge fixing of $v_m$. This $O(n)$
freedom can be used  to fix the remaining redundancies in
$a$ in any number of ways, for instance
to diagonalize any one of the four $\sigma_m$ components of the submatrix $a'$
(we use a different prescription in Sec.~8 below).

Since $O(n)$ has $\hf n(n-1)$ generators, we have succeeded, in principle,
in reducing the number
of independent scalar parameters in $a_{\lambda k}$ 
to $8n$ continuous degrees of freedom, as expected. But we are still
not done:  there may yet be a residual
\it discrete \rm degeneracy that needs to be modded out \Oone. To understand
this point,  imagine that the elements of $a$ are expressed
in some definite way in terms of $8n$ unconstrained
parameters, labeled  $\{ X_{i}\}$ with $i=1,\ldots, 8n$.
It may still happen that two different points in the parameter space, 
 $\{ X_{i}\}$ and $\{ \tilde X_{i} \}$,  correspond to
equivalent field configurations. From Eq.~\trans, we see that
 this occurs if there exists an $R\in O(n)$ such that
\eqn\condition{ \vW(\{X_{i}\})\cdot R\ =\ \vW(\{\tilde X_{i}\}) 
\qquad\hbox{and}\qquad R^{T}{a'}(\{X_{i}\})R
\ =\ {a'}(\{\tilde X_{i}\}) \ ,}
where $\vW=(w_{1},\ldots ,w_{n})$. 
In order to obtain the physical moduli space of inequivalent
instantons it is necessary to identify all such points. 
As we will see below, 
the degeneracy corresponds to the action of a discrete
symmetry group $G_{n}$ on the ADHM parameter space and the
required identification can be made by taking a quotient in the usual
way. The symmetry factor or ``statistical weight'' ${\cal S}_{n}$ 
of the parametrization counts the number of redundant
copies of each configuration. Hence ${\cal S}_{n}$ is given by the
number of solutions of Eq.~\condition\ which is just the order
of the symmetry group $G_{n}$. 
Knowledge of ${\cal S}_n$ is necessary in order properly to normalize
the integration measure over the moduli \Oone.

This concludes our overview of the ADHM construction of multi-instantons.
The reader may wish to verify that the single-instanton expressions
\adhmsing-\adhmreg\ follow from this construction for $n=1$. Fortunately in
what follows we will not actually need the explicit expressions for
multi-instantons, which are quite cumbersome even for $n=2.$ With
judicious use of integrations by parts, it will suffice to know only the
asymptotic behavior of the configurations as $|x|\rightarrow\infty.$
We therefore list for later use the asymptotic behavior of
several key ADHM quantities:
\eqna\asymadhm
$$\eqalignno{\Delta\ &\rightarrow\ bx\ ,&\asymadhm a
\cr f_{kl}\ &\rightarrow\ {1\over|x|^2}\,\delta_{kl}\ ,&\asymadhm b
\cr U_k\ &\rightarrow\ -{1\over|x|^2}\,x\,\wbar_k\,U_0\ ,&\asymadhm c
\cr U_0\ &\rightarrow\ \sigma_0\ ,&\asymadhm d}$$
where Eq.~\asymadhm{d} assumes the gauge choice \Uzerogauge.

\newsec{Multi-Instantons and Super-Multi-Instantons with Adjoint 
Matter}

This Section contains our principal findings for general winding number $n.$
In Sec.~7.1 we review the construction of adjoint fermion zero modes in
the general ADHM background. Sections 7.2-7.4 contain new results.
In Secs.~7.2-7.3 we construct the adjoint Higgs, first in the absence
of fermions, then in the presence of a Yukawa source term as in the $N=2$
\susic\ theory. A common feature of Secs.~7.1-7.3 is that, with the
correct ans$\ddot{\rm a}$tze, the solution to a linear differential equation
is mapped onto the solution of a linear finite matrix equation.
In Sec.~7.4
we construct the multi-instanton action in both the bosonic and the
\susic\ cases. Finally in Sec.~7.5
we discuss the formulation of the multi-instanton measure.

\subsec{Adjoint fermion zero modes}

As in the one-instanton case,  the classical gaugino and Higgsino fields
are the zero modes for $\Dbarslash$ (in contrast, $\Dslash$ is still
invertible). The most general such
solution  was found by Corrigan et al.~using the 
method of tensor products \refs{\CGTone,\CGone,\CWSone}.
In this approach, the required adjoint fermion zero
modes can be constructed from the simpler expressions for zero modes
in the fundamental representation of the gauge group. 
Explicitly for the gaugino we have 
(suppressing the ADHM indices but retaining the $SU(2)$
indices for clarity):
\eqn\lambdazm{(\lambda_\alpha)^\dbeta{}_\dgamma\ =\ 
\Ubar^{\dbeta\gamma}\M_\gamma f\,\bbar\, U_{\alpha\dgamma}\ -\
\Ubar^\dbeta{}_\alpha \,bf\M^{\gamma T}U_{\gamma\dgamma}\ ,}
where $\M_\gamma$ is an $(n+1)\times n$-dimensional matrix of constant
Grassmann spinors. From Eq.~\lambdazm\ we calculate (using some of
the tricks of Sec.~7.2 below):
\eqn\zmid{\Dbarslash^{\dalpha\alpha}
(\lambda_\alpha)^\dbeta{}_\dgamma\ =\ 
2\Ubar^{\dbeta\alpha}\,bf\big(\Deltabar^{\dalpha\gamma}\M_\gamma+
\M^{\gamma T}\Delta_\gamma{}^\dalpha\big)f\,\bbar\, U_{\alpha\dgamma}\ .}
Hence the condition for a gaugino zero mode \refs{\CGTone,\CGone}
is the following two sets of
linear constraints on $\M_\gamma$ which ensure that the right-hand side
vanishes (expanding $\Delta(x)$ as $a+bx$):
\eqn\zmconsone{\abar^{\dalpha\gamma}\M_\gamma\ =\ -\M^{\gamma T}a_\gamma
{}^\dalpha} and \eqn\zmconstwo{\bbar_\alpha{}^\gamma\M_\gamma\ =\ \M^{\gamma 
T}
b_{\gamma\alpha}\ .}
Counting the number of degrees of freedom, one finds $2n(n+1)$ individual
Grassmann numbers in the matrix $\M_\gamma$, subject to $n(n-1)$ 
constraints from 
each of Eqs.~\zmconsone\ and \zmconstwo, for a net of $4n$ gaugino
zero modes. Since in the $N=2$ theory
these are reduplicated in the Higgsino $\psi$ as well
(to which we associate the matrix $\N_\gamma$) we have $8n$ linearly 
independent modes in all, the same as for the bosons. 
The corresponding entries in the
matrices $\M_\gamma$ and $\N_\gamma$ are Grassmann collective
coordinates which must be integrated over just like their counterparts
$\{\xi,\xi',\etabar,\etabar'\}$  in the one-instanton sector. 

A crucial distinction among these modes should be made
 between the lifted modes,
on which the superinstanton action $S_\inst$ depends, and the unbroken
modes which do not appear in $S_\inst.$ Thus, in the 1-instanton sector,
we saw in Sec.~4.3 that the four superconformal modes $\etabar_\dalpha$
and $\etabar'_\dalpha$ are lifted, while the four \susic\ modes
$\xi_\alpha$ and $\xi'_\alpha$ are exact. 
We will soon find that this pattern is extended for general $n$ in the
following way.
There are always precisely four unlifted modes, corresponding to
a single spinor degree of freedom in each of $\M_\gamma$ and $\N_\gamma$: 
the one proportional to $\xi_{1,2}\,\sigma^{mn}\uv_{mn}\,,$ swept out by
the exact $N=2$ \susy\ generators $Q_1$ and $Q_2,$ respectively
(see Eq.~(\susytwodef $b,f$)$\,$). Comparing the expressions \vmnagain\ and
\lambdazm, we see that these are the zero modes for which
\eqn\globalsusy{\M_\gamma\ \propto\ b\quad\hbox{and}\quad\N_\gamma\
\propto\ b\ ,}
in harmony with the global translational mode \globaltrans. In this
instance the constraints \zmconsone-\zmconstwo\ simply boil down to the
bosonic constraints \crucial{b,c}.  The remaining $8n-4$ modes are
lifted by the Yukawa interactions. Among the lifted modes are the 
superconformal modes \sconfrewrite, discussed at the end of Appendix
C.1; in contrast to these, the remainder of the lifted modes do not
correspond to (approximate) Lagrangian symmetries.

This mode counting is quite different from the case of $N=1$ \susic\
Yang-Mills theory coupled to fundamental Higgs, where the number of
unbroken modes, rather than remaining constant, grows linearly
with $n$ \ADSone. Therefore, in those models, the different winding
number sectors do not interfere (e.g., the fermion mass term is
a pure 1-instanton effect), in contrast to the $N=2$ model at hand,
where correlators receive contributions simultaneously from 
all $n$, including $n=0.$

\subsec{The classical adjoint Higgs field}

Here we construct the classical adjoint Higgs field $A_\cl$ 
 in the general  multi-instanton background, and with the
fermions turned off. 
We assume that the adjoint Higgs is of the following general form:
\eqn\higgsansatz{i A_\cl{}^\dalpha{}_\dbeta\ =\ 
\Ubar_\lambda^{\dalpha\alpha}\,\A(x)_{\lambda\kappa\,\alpha}{}^\beta\,
U_{\kappa\beta\dbeta}}
where the $(n+1)\times(n+1)$-dimensional
matrix $\A_{\lambda\kappa}$ is to be determined. 
Note that $\A_{\lambda\kappa}$ is not strictly speaking a quaternion;
rather, it is a quaternion multiplied by a complex phase, namely that
carried by the vev $\vhiggs.$ Tracelessness of $A_\cl$
implies $\bar\A = -\A$ in the case of real vev; more generally
$\bar\A=-\A\cdot(\vbarhiggs/\vhiggs).$ 

{}From Eqs.~\adhmansatz\ and \projdef\ one calculates the commutator
(dropping indices from now on):
\eqn\commute{[\,v_n\,,\,iA_\cl\,]\ =\ -\partial_n\Ubar\,\P\,\A U
-\Ubar\A\,\P\,\partial_nU}
so that, with Eqs.~\Deltadef-\perpspace, 
\eqn\covderiv{\D_n\,iA_\cl\ \equiv\ \partial_n\,iA_\cl+
[\,v_n\,,\,iA_\cl\,]\ =\ 
-\Ubar\partial_n\Delta f\Deltabar\A U-\Ubar\A\Delta f\partial_n\Deltabar\,U
+\Ubar\partial_n\A\, U\ .}
This expression can be further differentiated with the help of
\eqn\fid{\partial_nf\ =\ -f\,\partial_n(\Deltabar\Delta)\,f
\ =\ -f\,(\sigmabar_n\bbar\Delta+\Deltabar b\sigma_n)\,f\ .}
A straightforward calculation gives
\eqn\kgadhm{\eqalign{\D^2\,iA_\cl\ =\ 
&4\Ubar\,\big\{\,bf\,\bbar\,,\,\A(x)\,\big\}\,U
-4\Ubar b f\cdot\trtwo\Deltabar\A(x)\Delta\cdot f\,\bbar U
\cr&+\Ubar\,\partial^2\A(x)\,U
-2\Ubar\, bf\sigma_n\Deltabar\,\partial^n\A(x)\,U
-2\Ubar\partial^n\A(x)\Delta\sigmabar_nf\,\bbar U\ .}}
In obtaining this relatively simple answer we have exploited, in addition
to the usual Pauli matrix identity \wessbagger\ $\sigma^n_{\alpha\dalpha}
\sigmabar_n^{\dbeta\beta}=-2\delta_\alpha{}^\beta\delta_\dalpha{}^\dbeta,$
the fact that
\eqn\newfact{\sigma^n\,\Deltabar b\,\sigma_n\ =\
2(\bbar\Delta)^T\ =\ 2\bbar\Delta\ .}
The first equality in \newfact\ follows from $\sigma^n\sigmabar_m\sigma_n=
2\sigma_m$ and the second from the ADHM constraints \crucial{}.

We wish to solve $\D^2\,A_\cl=0$ together with the vev boundary conditions
\eqn\vevbc{iA_\cl{}^\dalpha{}_\dbeta\ \buildrel
|x|\rightarrow\infty \over \longrightarrow\
\textstyle {i\over2}\,\vhiggs^a\,\tau^{a\dalpha}{}_\dbeta\ .}
As we stressed in the Introduction, 
our general strategy in solving this type of ADHM equation
is to seek inspiration whenever possible from 
the 1-instanton sector. In that instance we need only compare
Eqs.~\sing{c} and \adhmsing\ to see the answer. We find that $\A(x)$
is in fact a constant, namely
\eqn\littleblock{\A(x)\ =\
\pmatrix{\vhiggsa&0\cr0&0}\ ,}
where
\eqn\vevbcagain{{\vhiggsa}_\alpha{}^\beta
\ =\ \textstyle{i\over2}\,\vhiggs^a\,\tau^a{}_\alpha{}^\beta\ .}
Extrapolating to multi-instantons, we will guess that $\A(x)$
remains  a constant. 
{}From the asymptotic behavior \asymadhm{c,d} for $U_\lambda$, we 
then draw two important conclusions. First, we see that
the top-left entry, $\vhiggsa$,
 must by itself account for the boundary conditions \vevbc;
thus Eq.~\vevbcagain\ continues to hold for multi-instantons as well.
Second, if we assume that the \it next\rm-leading
behavior of $A_\cl$ goes like $1/|x|^2$ rather than $1/|x|,$ again as in
 the 1-instanton sector, then the
$(\lambda=0,\kappa\neq0)$ and $(\lambda\neq0,\kappa=0)$ 
elements of $\A$ should
vanish. In short,  our ansatz is
\eqn\blockdiag{\A_{\kappa\lambda}\ =\
\pmatrix{\vhiggsa&0&\cdots&0 \cr 0&{}&{}&{}\cr
\vdots&{}&\A'_{kl}&{}\cr 0&{}&{}&{}}\quad,\ \quad
\partial_n\A\,=\,0\ .}
Consequently, Eq.~\kgadhm\  collapses to
\eqn\newkgadhm{\D^2\,iA_\cl\ =\ 
4\Ubar'\,\big\{\,f\,,\,\A'\,\big\}\,U'
-4\Ubar'  f\cdot\trtwo\Deltabar\A\Delta\cdot f U'\ ,}
where $U_l'$ is the truncated vector obtained by lopping off the $\lambda=0$
component from $U_\lambda.$

To make further progress, it is convenient to denote by $\Delta_{kl}'(x)
=a_{kl}'+\delta_{kl}\cdot x$ the lower 
$n\times n$ submatrix of $\Delta_{\kappa l}(x).$
We also  define the three $n\times n$ matrices $\Lambda,$ $\widetilde W$,
and $W=\trtwo\widetilde W$ by their  matrix elements
\eqn\newmatdef{\Lambda_{kl}\,=\,\wbar_k\vhiggsa w_l-
\wbar_l\vhiggsa w_k\ ,\quad
\widetilde W_{kl}\,=\,\wbar_kw_l\ ,\quad
W_{kl}\,=\,\wbar_kw_l+\wbar_lw_k\ .}
We then rewrite
\eqn\trrewrite{\eqalign{\trtwo\Deltabar\A\Delta\ &=\,
\Lambda+\trtwo\Deltabar'\A'\Delta'\cr
&=\, \Lambda+\hf\trtwo\Big(
[\,\Deltabar'\,,\,\A'\,]\Delta'-\Deltabar'[\,\Delta'\,,\,\A'\,]
+\{\,\A'\,,\,\Deltabar'\Delta'\,\} \Big)
\cr
&=\, \Lambda+\hf\trtwo\Big(
[\,\Deltabar'\,,\,\A'\,]\Delta'-\Deltabar'[\,\Delta'\,,\,\A'\,]
-\{\,\A'\,,\,\widetilde W\,\}
+\{\,\A'\,,\,\finv\,\} \Big)}}
where the final equality follows from Eq.~\Deltaid.

Notice that 
the last term in Eq.~\trrewrite, namely $\hf\{\,\trtwo\A'\,,\,\finv\,\},$
gives a contribution that is naturally combined with the anticommutator
in Eq.~\newkgadhm, for a total of
$4\Ubar'\,\{\,f\,,\,\A'-\hf\trtwo\A'\,\}\,U'$. With the obvious
further ansatz
\eqn\Asinglet{\A'_{kl\,\alpha}{}^\beta\ =\ \A'_{kl}\,\delta_\alpha{}^\beta}
this vanishes, leaving only the terms containing two factors of $f$:
\eqn\kgadhmagain{\D^2\,iA_\cl\ =\ 
-4\Ubar'  f\cdot
\Big(\ \Lambda-\hf\{\,\A'\,,\,W\,\}+\hf\trtwo\big(\,
[\,\abar'\,,\,\A'\,]a'-\abar'[\,a'\,,\,\A'\,]\,\big)\ \Big)\cdot f U'\ .}
Thanks to \Asinglet\ the $\delta_{kl}\cdot
x$ part of $\Delta'_{kl}(x)$ has canceled 
out of the commutator terms. In fact, with the $x$-dependence  gone, the
entire $n\times n$ matrix in big parentheses can now sensibly be set to zero.
With $a'_{\alpha\dalpha}=
a'_m\sigma^m_{\alpha\dalpha}$ we arrive, finally, at the defining equation
for the classical adjoint Higgs:
\eqn\defineAcl{-[\,a^{\prime m}\,,\,[\,a'_m\,,\,\A'\,]\,]\ +\
\hf\{\,\A'\,,\,W\,\}\ =\ \Lambda\ .}
{}From Eqs.~\newacond, \blockdiag, \newmatdef\ and \Asinglet, together with
the requirement $\bar\A=-\A\cdot(\vbarhiggs/\vhiggs),$ we see that
 both $\A'_{kl}$ and $\Lambda_{kl}$ are antisymmetric while
$W_{kl}$ and $a'_{kl}$ are symmetric; therefore
both sides of this matrix equation are consistent. 
For the $n$-instanton sector,
 Eq.~\defineAcl\ defines a set of $\hf n(n-1)$ coupled linear
inhomogeneous equations for the $\hf n(n-1)$
independent entries of
 $\A',$ in terms of the constrained $(n+1)\times n$-dimensional
ADHM matrix $a.$ We note for future reference \dkmtwo\ that the
$\hf n(n+1)\times\hf n(n+1)$ dimensional linear operator on the space
of antisymmetric matrices defined by the left-hand side of \defineAcl\
is actually self-adjoint on this space, as the reader can easily check
in the obvious basis.

For an alternative route to Eq.~\defineAcl, see Appendix~C.

\subsec{The adjoint Higgs in the presence of a Yukawa source}

Next we solve the more challenging equation \eulermatter{a}, for the
adjoint Higgs in the presence of an adjoint
 fermion bilinear source term, subject
once again to the vev boundary conditions \vevbc. 
In terms of $\M_\gamma$ and $\N_\gamma$ the right-hand side of
Eq.~\eulermatter{a} is expanded as follows:
\eqn\rhsema{\eqalign{-\sqrtwo i \,\Ubar^{\dbeta\gamma}\,
\Big(\ &\N_\gamma f\,\bbar\,\P^{\delta\alpha}\,\M_\alpha f\,\bbar
\ +\ bf\N^{\alpha T}\,\P_{\alpha\gamma}\,bf\M^{\delta T}
\cr +\ &\N_\gamma f\,\bbar\cdot\trtwo\,\P\,\cdot bf\M^{\delta T}
\ -\ \delta_\gamma{}^\delta bf\N^{\alpha T}\,\P_\alpha{}^\beta\,\M_\beta
f\,\bbar\ \Big)\,U_{\delta\dgamma}
\cr-\ \big(\,&\M\ \rightleftharpoons\ \N\,\big)\ .}}
We can ignore the auxiliary field D at this order, as in the 1-instanton
case.
To solve for the Higgs we exploit the linearity of the 
Klein-Gordon equation,
and decompose $A$ into $A_\cl+A_f$. As constructed in the
preceding Section, $A_\cl$  solves the homogeneous equation and soaks up the
boundary conditions \vevbc. This leaves $A_f$ (the subscript stands for
``fermionic'') to account for the Yukawa source \rhsema, while approaching
zero as $|x|\rightarrow\infty.$ 

Unlike the 1-instanton case of Sec.~4, here we cannot rely on a
``sweeping-out'' procedure to generate $A_f$ automatically. The reason
is that for $n>1,$ the superconformal group will only produce a fixed
subset of the adjoint fermion zero modes; most of these zero modes are
not associated with (approximate) Lagrangian symmetries. Nevertheless,
an intelligent first guess for
$A_f$ may be intuited once again from the 1$\hbox{-}$instanton sector.
There, $A_f$ is the fermion bilinear piece of Eq.~\losuper{b}
proportional to $\xi'(x)\sigma^{mn}\xi(x)\uv^\cl_{mn}$,
where $\xi(x)$ and $\xi'(x)$ parametrize the supersymmetric and
superconformal modes of, respectively,  the gaugino and the Higgsino.
This expression may be regarded  either  as the
Higgsino collective coordinates dotted into the most general gaugino zero
mode, or, symmetrically, as 
the gaugino collective coordinates dotted into the most general Higgsino zero
mode. Extrapolating to the $n$-instanton case at hand, and dotting
$\N^\alpha$ into the most general gaugino zero mode \lambdazm, we are led 
immediately to the symmetric ansatz
\eqn\Aftry{i(A_f)^\dbeta{}_\dgamma\  \buildrel ? \over = \ {1\over2\sqrtwo}\
\ \Ubar^{\dbeta\alpha}\,\Big(\,\N_\alpha\,f\,\M^{\beta T}\ -\ 
\M_\alpha\,f\,\N^{\beta T}\,\Big)\,U_{\beta\dgamma}\ ,}
where the constant in front comes from a careful comparison of
normalizations. 

In actuality, for $n>1$ this guess for $A_f$ is
almost, but not  quite, correct, as the reader can check by plugging
it into Eq.~\kgadhm. Specifically, 
it  accounts for everything in
Eq.~\rhsema\ except the last term, where it gives $\P-1$ rather than
$\P$ in the middle of the expression. Taking advantage of the linearity of
Eq.~\eulermatter{a} for a second time, one further decomposes
\eqn\Aftryagain{i(A_f)^\dbeta{}_\dgamma\   = \ {1\over2\sqrtwo}\
\ \Ubar^{\dbeta\alpha}\,\Big(\,\N_\alpha\,f\,\M^{\beta T}\ -\ 
\M_\alpha\,f\,\N^{\beta T}\,\Big)\,U_{\beta\dgamma}\ +\ i
(A'_f)^\dbeta{}_\dgamma\ .}
By design, $A'_f$ accounts for the missing piece of the last term:
\eqn\Afprimedef{\D^2\,(A_f')^\dbeta{}_\dgamma\ =\ -4 i\,
\Ubar^{\dbeta\alpha}\,bf\,\Lambda_f\,
f\,\bbar\, U_{\alpha\dgamma}\ ,}
with
\eqn\lambdafdef{ \Lambda_f\, =\,-\Lambda_f^T\,=\,
 -{1\over2\sqrtwo}\,
\big(\,\N^{\beta T}\M_\beta-\M^{\beta T}\N_\beta\,\big)\  .}
Fortunately, this equation is easily solved by  the methods of Sec.~7.2.
 Making the ansatz
\eqn\Afprimeansatz{iA_f'\ =\ \Ubar\cdot
\pmatrix{0&\cdots&0\cr\vdots&\A_f'&{}\cr0&{}&{}}\cdot U}
where 
\eqn\Aprimeconds{\partial_n\,\A'_f\,=\,0\ ,\quad
(\A'_f){}^{}_{kl\,\alpha}{}^\beta\,=\,(\A'_f){}^{}_{kl}\,
\delta_\alpha{}^\beta \ ,\quad(\A'_f){}^{}_{kl}=-(\A'_f){}^{}_{lk}}
as before, one regains Eq.~\kgadhmagain, where now
$\A'\rightarrow\A'_f$ and $\Lambda\rightarrow\Lambda_f$. The solution
follows by direct analogy with Eq.~\defineAcl,
 and is given by the antisymmetric matrix equation
\eqn\defineAprimef{-[\,a^{\prime m}\,,\,[\,a'_m\,,\,\A_f'\,]\,]\ +\
\hf\{\,\A_f'\,,\,W\,\}\ =\ \Lambda_f\ .}
 Like Eq.~\defineAcl,
this defines a set of $\hf n(n-1)$ linear inhomogeneous equations for the
$\hf n(n-1)$ independent scalars in $\A_f',$ completing our task.

\subsec{Multi-instanton and super-multi-instanton action}

We now construct the supersymmetric multi-instanton action, through
order $g^0$. The pure Yang-Mills contribution is, of course,
$S_\cl=8n\pi^2/g^2$. An instant way to derive this fact, or equivalently
that the winding number is $n$, is to think of the field strength 
\vmnagain\ as a translational vector zero mode, i.e.~Eq.~(B.1) with
$C^{\sst(m)}=g^{-1}b\sigma_m,$ and then to use the integration formula
(B.5). (Alternatively one can exploit the interesting identity 
\refs{\Jack,\Oone} $\trtwo v_{mn}v^{mn}=\partial^4\log\det f^{-1}.$)

At the next order in $g^2|\vhiggs|^2$ one needs
to evaluate the surface integral shown in Eq.~\instaction.
In the long-distance
limit we can simply replace $A^\dagger$ in that expression by
$\bar\vhiggsa$, as in the 1-instanton case.
With the help of Eqs.~\asymadhm{} and \covderiv\ we 
 calculate  the gradient of the bosonic and fermion-bilinear
pieces of the Higgs, respectively:\foot{We can ignore pre- and
post-multiplication by $\sigmabar_0$ and $\sigma_0$ which merely
interchanges dotted and undotted indices. Note also that unlike for the
bosons, the order in $g$ of a fermionic contribution to the action
is convention-dependent; it is a property of Grassmannian integrals that
any rescaling of the fermionic action can be compensated by a change
in the measure.}
\eqna\limiting
$$\eqalignno{
\hat x_m\,\D^m A_\cl \ &\buildrel {|x|\rightarrow\infty}\over
\longrightarrow\ 
{2\over|x|^3}\,\big(\,\vhiggsa{}_\alpha{}^\beta
\sum_k|w_k|^2\,-\,w_{k\alpha\dalpha}\,\A'_{kl}\,
\wbar_l^{\dalpha\beta}\,\big)\ ,&\limiting a
\cr
\hat x_m\,\D^m A_f \ &\buildrel {|x|\rightarrow\infty}\over
\longrightarrow\ 
-{1\over|x|^3\sqrtwo}\,\big(\,\nu_{k\alpha}\mu^\beta_{k}\,-\,
\mu_{k\alpha}\nu^\beta_{k}\,\big)\,\cr&\qquad\quad
-\,{2\over|x|^3}\,w_{k\alpha\dalpha}
\,(\A'_f)^{}_{kl}\,
\wbar^{\dalpha\beta}_l\ ,&\limiting b}$$
where $\mu_k$ and $\nu_k$ denote the top-row elements
\eqn\MNlabel{\mu_k^\alpha\ =\ \M_{0k}^\alpha\ ,\quad
\nu_k^\alpha\ =\ \N_{0k}^\alpha\ .}
Substituting Eqs.~\limiting{} into Eq.~\instaction\ and performing the traces
gives, finally,
\eqn\sinstfinal{\eqalign{S_{\rm inst}\ &\equiv\ S_\cl+
S_\higgs+S_\yuk\cr
&=\ {8n\pi^2\over g^2}\ +\ 16\pi^2|\vhiggsa|^2\sum_k|w_k|^2\ 
-\ 8\pi^2\big(\wbar_l\bar\vhiggsa w_k-\wbar_k\bar\vhiggsa w_l\big)\,\A'_{kl}
\cr&\quad 
+\ 4\sqrtwo\pi^2\,\mu_k^\alpha\,\bar\vhiggsa{}_\alpha{}^\beta\,\nu_{k\beta}
-\ 8\pi^2\big(\wbar_l\bar\vhiggsa w_k-\wbar_k\bar
\vhiggsa w_l\big)\,(\A'_f){}^{}_{kl}
\ .}}
We remind the reader where the various threads of this equation may be
located: Eqs.~\bcanonical, \lambdazm, \vevbcagain, \newmatdef, \defineAcl,
 \Aftryagain, \lambdafdef, \defineAprimef, and \MNlabel.

The expression \sinstfinal\ 
constitutes the exact classical interaction between an
arbitrary number of superinstantons in $N=2$ \susic\ Yang-Mills
theory and is our main result in the general case; eliminating the
last line gives the analogous result in the purely bosonic version
of the theory. Some comments about the parts and the whole:

\bf1\rm.
Beyond its role in the action, the quantity $S_\higgs$ also governs the
asymptotics of various component fields in the problem. For example,
we discover that the 1-instanton relation \delAbarlong\ between
$\delta\Abar$ and $S_\higgs$ continues to hold in the multi-instanton
case, even though $S_\higgs$ is much more complicated. On reflection
this is no surprise, for Eqs.~\delAbarlong\ and 
\limiting{a} are essentially equivalent statements. Similarly for
the antifermions specifically in the background of the exact \susic\
mode, which are proportional to $\Dslash\Abar$ (see Eqs.~(\susytwodef{\it
c,i})$\,$); again  the 1-instanton expressions \psibarlong-\lambdabarlong\
generalize immediately to $n\ge1.$  We will need these expressions 
once more in the following Section.

\bf2\rm.
As for $S_\yuk,$ the reader can check that the only Grassmann modes
not lifted by this expression are the exact \susy\ modes \globalsusy.
Thanks to the constraint \zmconstwo, these modes cancel out of
$\Lambda_f,$ and therefore do not appear in $\A_f'.$

\bf3\rm. In Ref.~\dkmtwo\ we verify that the sum $S_\higgs+S_\yuk$ as
given above is in fact a \susic\ invariant, as it must be.

\subsec{Multi-instanton measure}

To proceed to a complete calculation of the multi-instanton
contributions to the prepotential $\cal F$, we require the measure for
integration over the moduli. 
The collective coordinate integration measure for $n$ ADHM instantons 
in $N=2$ \susic\ Yang-Mills theory is obtained in the usual way 
by changing variables in the functional
integral from the fields to the moduli \refs{\tHooft,\Amati}. 
Listing the $8n$ unconstrained bosonic and fermionic coordinates as $X_{i}$
and $\xi_{i}$, respectively, we have 
\eqn\measureone{
\int\, d\mu_n\ =\ {1\over{\cal S}_n}\int \,
\left(\prod_{i=1}^{8n} dX_{i}d\xi_{i}\right) 
\big(\,J_\bose/J_\fermi\,\big)^{1/2}\,\exp(-S_\inst)
 }
where $J_\bose$ ($J_\fermi$) is the Jacobian for the change of variables
for the bosonic (fermionic) degrees of freedom. 
A key simplification in a supersymmetric theory is that there is no
additional small-fluctuations 't Hooft determinant to be calculated,
as the positive frequency bosonic and fermionic excitations cancel 
identically. As we discussed above,
any unconstrained parametrization of the $n$-instanton solution will,
in general, contain several redundant copies of each field
configuration. Hence, to obtain the correct normalization of the
measure,  we must divide out the relevant symmetry factor 
${\cal S}_{n}$. 

The fermionic Jacobian, $J_\fermi$, is simply the determinant of the
normalization matrix of the fermion zero modes. 
In Appendix B, we calculate this matrix in
terms of the (constrained) bosonic parameters of the  ADHM solution
for general $n$, and evaluate its determinant for $n=2$. 
The bosonic Jacobian has been studied in detail by
Osborn \Oone.  $J_\bose$ is much harder to evaluate than
$J_\fermi$, because
the corresponding variations of the collective coordinates are not
simply equal to the physical bosonic zero modes
but differ from them by  transformations of the type 
\inserta-\insertd. In particular an 
explicit expression for $J_\bose$ cannot be obtained without first
specifying an unconstrained parametrization of the solution. 
Given any such parametrization, the required measure can then be
constructed using the methods developed by Osborn and the
corresponding multi-instanton contribution to the prepotential can, in
principle, be 
written as a finite-dimensional integral. Unfortunately, as mentioned
above, the constraints have only been solved for $n\leq 3$ so we
cannot make immediate progress in the general case. 

It can easily be checked that applying our general results to the case
$n=1$ reproduces all the known results for the action, the measure and
the modes in the one-instanton sector. One result we will need below
is the correctly normalized measure $d\mu_{1}$ for a single ADHM
instanton. The bosonic and fermionic collective coordinates of the
$n=1$ ADHM super-instanton are contained in the three $2\times 1$
matrices of unconstrained parameters:
\eqn\oneI{a \,=\, \left({w \atop X} \right) \ ,\qquad
{\cal M}_{\gamma}\, =\,\left(
{\mu_{\gamma} \atop M_{\gamma}}\right)\ ,\qquad
{\cal N}_{\gamma}\,=\,\left(
{ \nu_{\gamma} \atop N_{\gamma}}\right)\ .}
These coordinates are related in a simple way to the parameters of the single
instanton appearing in the measure \measure: $X=x_{0}$, $|w|=\rho$,
$M=4\xi$, $N=4\xi'$, $\mu=4w\bar{\eta}$, $\nu=4w\bar{\eta}'$.    
In these variables the correctly normalized measure 
\measure\ can be rewritten as
\eqn\measureone{
\int d\mu_{1}\ =\ 2^{7}\pi^{-4}\Lambda^{4}\int\, 
d^{4}Xd^{4}wd^{2}Md^{2}Nd^{2}\mu d^{2}\nu\,\exp(-S_\higgs-S_\yuk)
 \ .}

In the following Section  we focus on the case $n=2$.

\newsec{The Uses of $N=2$, $n=2$ Superinstantons}

\subsec{The collective coordinates,  the measure, and the action}

We now specialize to the 2-instanton sector. The $16$ gauge, $8$ gaugino and
$8$ Higgsino collective coordinates live, respectively, in the following
matrices:
\eqna\twoinstmats
$$\eqalignno{
a\ &=\ \pmatrix{w_1&w_2\cr x_0+a_3&a_1\cr a_1&x_0-a_3}\ \ ,&\twoinstmats a
\cr
\M_\gamma\ &=\ \pmatrix{\mu_{1\gamma}&\mu_{2\gamma}\cr
4\xi_{\gamma}+\M_{3\gamma}&\M_{1\gamma}\cr
\M_{1\gamma}&4\xi_{\gamma}-\M_{3\gamma}}&\twoinstmats b
 \cr
\N_\gamma\ &=\ \pmatrix{\nu_{1\gamma}&\nu_{2\gamma}\cr
4\xi'_{\gamma}+\N_{3\gamma}&\N_{1\gamma}\cr
\N_{1\gamma}&4\xi'_{\gamma}-\N_{3\gamma}}\ .&\twoinstmats c}$$
As in the 1-instanton case 
the modes $x_0$, $\xi_{\gamma}$ and $\xi'_{\gamma}$ represent, respectively,
global translations and global $N=2$ supersymmetries $Q_1$ and $Q_2$;
see Eqs.~\globaltrans\ and \globalsusy.  The lower 
$2\times2$ sub-block of each of these matrices is symmetric, as forced
by the constraints \newacond\ and \zmconstwo. The remaining constraints
\crucial{a} and \zmconsone\ may be used to eliminate $a_1,$ $\M_1$ and
$\N_1$:
\eqn\aonedef{ a_1\ =\ {1\over4|a_3|^2}\,a_3(\wbar_2w_1-\wbar_1w_2+\Sigma)\ ,}
\eqn\Monedef{\M_1\ =\ {1\over2|a_3|^2}\,a_3\,\big(\,2\abar_1\M_3+\wbar_2\mu_1
-\wbar_1\mu_2\,\big)\ ,}
and
\eqn\Nonedef{\N_1\ =\ {1\over2|a_3|^2}\,a_3\,\big(\,2\abar_1\N_3+\wbar_2\nu_1
-\wbar_1\nu_2\,\big)\ .}
Notice that the ADHM constraints have only determined the four quaternionic
components of $a_1$ up to a new ``seventeenth collective coordinate'' 
$\Sigma,$
which apart from the requirement $\Sigma^\dalpha{}_\dbeta\propto
\delta^\dalpha{}_\dbeta$ is completely unrestricted. 
This is precisely consistent with our counting in Sec.~6.2; as discussed
there, this extra degree of freedom may be eliminated in a variety of
ways by invoking the left-over $O(2)$ symmetry \trans.
%
 In the following
we will use this $O(2)$ freedom to fix $\Sigma\equiv 0$ 
which simplifies the algebra enormously. In Appendix D we find that
the discrete symmetry group of this parametrization is the dihedral group
$D_{8}$ and hence the corresponding symmetry factor is ${\cal
S}_{2}=16$.   

Following Sec.~7.5, the 
$N=2,$ $n=2$ superinstanton measure is written as 
 \eqn\intmeas{\eqalign{
&{{1}\over {{\cal S}_{2}}}\int\,d^4x_0d^4a_3d^4w_1d^4w_2\times d^2\xi 
d^2\M_3d^2\mu_1
d^2\mu_2\times d^2\xi' d^2\N_3d^2\nu_1d^2\nu_2
\cr
&\quad\quad\times\ \exp(-S_{\rm inst})\,\big(\,J_\bose/J_\fermi\,\big)^{1/2}
\ .}}
It is helpful to have in mind a physical picture of these integration
variables \CWSone.
As discussed above the coordinates $x_{0}$, $\xi$ and $\xi'$ 
can always be thought of as 
the location of the center of the two-instanton configuration in
superspace. In contrast,
the other coordinates can only be given a definite interpretation in
a clustering limit where the two instanton solution is approximately
the linear superposition of two single instantons. For the 
choice $\Sigma\equiv0$, one such limit is $|a_{3}|\rightarrow\infty$. In this
limit the off-diagonal elements in the collective coordinate matrices
go to zero and it is straightforward to identify the corresponding  
one-instanton degrees of freedom. In particular, the combinations
$x_{(1)}=-(x_{0}+a_{3})$ and $x_{(2)}=-(x_{0}-a_{3})$ can be identified as
the centers of two well-separated singular gauge instantons. The scale
sizes of these instantons are given by the magnitudes of the
quaternions $w_{1}$ and $w_{2}$, $\rho_{(i)}=|w_{i}|$ for $i=1,2$, while
their $SU(2)$ iso-orientations (see Eq.~\adhmreg\ \it ff\rm.) 
are given by the corresponding unit-normalized
quaternions, $w_1/\rho_{(1)}$ and $w_2/\rho_{(2)}$. The relationship
between the $\mu_i$ and $\nu_i$ with the superconformal modes is also
straightforward; see Appendix C.1.

Next we discuss the integrand of Eq.~\intmeas.
It is convenient to define the four frequently occurring
combinations of the bosonic parameters from Eqs.~\vevbcagain\ and
\twoinstmats{}, as follows:
\eqna\paramsdef
$$\eqalignno{
L\ &=\ |w_1|^2+|w_2|^2\ ,&\paramsdef a
\cr
H\ &=\ L+4|a_1|^2+4|a_3|^2\ ,&\paramsdef b\cr
\Omega\ &=\ w_1\wbar_2-w_2\wbar_1\ ,&\paramsdef c\cr
\omega\ 
&=\ \hf\trtwo\Omega \vhiggsa\ =\ \wbar_2\vhiggsa w_1-\wbar_1\vhiggsa w_2
\ ,&\paramsdef d}$$
in terms of which the action \sinstfinal\ works out to
\eqn\newSdef{\eqalign{S_{\rm inst}\ &\equiv\ S_\cl+S_\higgs+S_\yuk\ \cr&=\
{16\pi^2\over g^2}\,+\,16\pi^2
\Big(\,L|\vhiggsa|^2\,-\,{|\omega|^2\over H}
\,\Big)\ \cr&\quad+\
4\sqrtwo\pi^2\Big(\,-\nu_k\bar\vhiggsa\mu_k\,+\,(\bar\omega/H)(\mu_1\nu_2-
\nu_1\mu_2+2\M_3\N_1-2\N_3\M_1)\, \Big)\ .}}

Remarkably, Osborn \Oone\ obtained an explicit expression for the 
bosonic collective coordinate Jacobian for an
arbitrary unconstrained parametrization of the two
instanton solution.\foot{The measure for the particular parametrization of
Jackiw, Nohl and Rebbi \JNRone\ had been reported earlier in 
Refs.~\refs{\GMOone,\MAone}. } In our notation, Osborn's result reads
\eqn\Jbosepropto{J_\bose^{1/2}\ \propto\ {H\over|a_3|^4}\,\Big|\,|a_3|^2
\,-\,|a_1|^2\,-\,\eighth{d\Sigma\over d\theta}\big|_{\theta=0}\,\Big|\ ,}
where the angle $\theta$ parametrizes the $O(2)$ symmetry \trans.
As for
the fermionic Jacobian, it is given by the determinant of the overlap
matrix of the fermion zero modes. We calculate this explicitly in
Appendix C and find:
\eqn\Jfermipropto{J_\fermi^{1/2}\ \propto\ {H^2\over|a_3|^4}\ .}
Putting the factors which occur in the measure together 
and specializing to  $\Sigma\equiv 0$, we obtain:
\eqn\Jratio
{{{1}\over {\cal S}_{2}}\big(\,J_\bose/J_\fermi\,\big)^{1/2}
\,\exp(-S_\cl)\ =\ \cJ\,
{\big|\,|a_3|^2-|a_1|^2\,\big|\over H}\ .}

The overall constant $\cJ$ in Eq.~\Jratio\ 
can easily be determined by noting that the Jacobians, being local
functionals of the fields, factorize into the
product of single-instanton Jacobians in the clustering limit 
$|a_{3}|\rightarrow\infty$. Taking account of the symmetry factors, this
implies the following relationship between the 
the correctly normalized measures, $d\mu_{n}$, 
for the cases $n=2$ (with this particular parametrization) and $n=1$:
\eqn\statmech{\int d\mu_{2}\ \longrightarrow\
 {{\cal S}^{2}_{1}\over{\cal S}_{2}}
\int\, d\mu^{(1)}_{1} \times d\mu^{(2)}_{1}  }
as $|a_{3}|\rightarrow \infty$. Here $d\mu^{(1)}_{1}$ and
$d\mu^{(2)}_{1}$ are the normalized measures for the collective 
coordinates of two well-separated single instantons. 
In Appendix D we note that the symmetry
factor for the single instanton measure is given by ${\cal S}_{1}=2$.
 Hence using ${\cal S}_{2}=16$ and Eq.~\measureone\
we extract the value
\eqn\cjequalscdots{\cJ\ =\ 2^{6}\pi^{-8}\Lambda^{8}\ ,} 
which completes the specification of the 2-instanton measure.

\subsec{$4$-fermi Green's functions, and a $2$-instanton
check on Seiberg and Witten}

In what follows we shall focus on the family of 4-fermi Green's functions
\familypoint\
that we examined earlier in the 1-instanton sector. The 28-fold integration
 proceeds step-wise, and more or less painlessly, as follows:

\bf1\rm. As in the 1-instanton case,
the four antifermions are needed to saturate the integration over the
exact \susic\ modes $d^2\xi\,d^2\xi'$ which do not otherwise appear
in the integrand. The remaining (lifted) Grassmann modes are then necessarily
saturated by pulling down appropriate powers of the action.
A straightforward calculation gives for these lifted modes:
\eqn\yukint{\eqalign{
&\int\,d^2\M_3d^2\mu_1d^2\mu_2d^2\N_3d^2\nu_1d^2\nu_2\ 
\exp(-S_\yuk)\cr&\ \ =
-\,\Big({16\sqrtwo \pi^6\bar\omega\over|a_3|^2\,H\,|\Omega|}
\Big)^2\,\Big(\ \big(\,-(\bar\vhiggsa)^2|\Omega|^2+\tau_1\bar\omega^2\,\big)^2
\ -\ \tau_2\bar\omega^2\big(\,(\bar\vhiggsa)^2|\Omega|^2+\bar
\omega^2\,\big)\ \Big)}}
where $\tau_1 = L/H$ and $ \tau_2 = (L^2-|\Omega|^2)/ H^2$.
Comparing this expression to its 1-instanton counterpart \sconfint\ reflects
the order-of-magnitude increase in complexity in the step between
the 1-instanton and 2-instanton sectors.
It is convenient to substitute
 $\bar\omega^2=|\omega|^2\cdot(\vbarhiggs/\vhiggs)$
and also
$(\bar\vhiggsa)^2=-|\vhiggsa|^2\cdot(\vbarhiggs/\vhiggs)=-\vbarhiggs^2/4.$

\bf2\rm. 
We now insert the $l+4$ fields from \familypoint, or more specifically
their asymptotics as given in Eqs.~\delAbarlong, \psibarlong\ and
\lambdabarlong, and perform
the trivial integration over the exact \susic\ modes $d^2\xi\,d^2\xi'.$
We are left with a purely bosonic integral. The integrand is greatly
simplified by the familiar trick of letting $S_\bose\rightarrow\lambda
S_\bose$ in the action, and representing the $l+4$ powers of $S_\bose$
coming from these equations 
as parametric derivatives with respect to $\lambda.$ (Alternatively,
if one is careful to maintain the distinction between $\vhiggs$ and
$\vbarhiggs,$ these field insertions may be obtained from the action
\newSdef\ by differentiation with respect to $\vhiggs$.)

\bf3\rm. Next we change integration variables from
the quaternions $\{a_3,w_1,w_2\}$ into the more natural
coordinates in the problem, $\{H,L,\Omega\}$. Using Eq.~\aonedef, and
fixing $\Sigma\equiv0$, we calculate
\eqn\covone{\eqalign{\int_{-\infty}^\infty d^4a_3\ 
{\big|\,|a_3|^2-|a_1|^2\,\big|\over |a_3|^4}\ &=\
\pi^2\int_0^\infty|a_3|^2\,d|a_3|^2\ 
{\big|\,|a_3|^2-|a_1|^2\,\big|\over |a_3|^4}
\cr&\longrightarrow\ 2\times{\pi^2\over4}\,\int_{L+2|\Omega|}^\infty\ dH}}
and likewise
\eqn\covtwo{\eqalign{\int_{-\infty}^\infty\,d^4w_1\,d^4w_2\ 
&\longrightarrow\ 2\times\int_0^\infty dL\ \int_{|\Omega|\le L}d^3\Omega\ 
\int_{L_-}^{L_+}\pi^2|w_1|^2\,d|w_1|^2\ \cr&\qquad\qquad\qquad\qquad
\times\ {1\over16|w_1|^2\,\sqrt{(L_+-|w_1|^2)(|w_1|^2-L_-)}}
\cr&=\ 
2\times{\pi^3\over16}\int_0^\infty dL\ \int_{|\Omega|\le L}d^3\Omega\ .}}
In Eq.~\covone\ the numerator and denominator of the integrand are supplied
by Eqs.~\Jratio\ and \yukint, respectively; in Eq.~\covtwo\ the 
(irrelevant) limits
of integration are $L_\pm=\hf(L\pm\sqrt{L^2-|\Omega|^2}\,)$; and in
both Eqs.~\covone\ and \covtwo\ the extra overall factors of 2 reflect
the fact that each of these changes of variables is a 2-to-1 mapping.

\bf4\rm. Next we rescale variables,
\eqn\omegarescale{\Omega'\,=\,\Omega/L\ ,\quad
H'\,=\,H/L\ ,\quad\ \omega'\,=\,\hf\trtwo\Omega'\vhiggsa\,,}
and switch to polar coordinates
\eqn\polarcoor{d^3\Omega'\ \longrightarrow\ 2\pi\int_{-1}^1d(\cos
\theta)\int_0^1|\Omega'|^2d|\Omega'|}
where the polar angle is defined by $|\omega'|=|\Omega'||\vhiggsa|
\cos\theta=\hf|\Omega'||\vhiggs|\cos\theta.$ With
 a bookkeeper's eye on the various factors of two as well as on the
distinction between $\vhiggs$ and $\vbarhiggs,$ we can then reexpress
the Green's function \familypoint\ as:
\eqn\Gfbecomes{\eqalign{\int d^4x_0& \int_0^1d|\Omega'|\,|\Omega'|^6
\int_{-1}^1d(\cos \theta)\,\cos^2\theta\int_{1+2|\Omega'|}^\infty
{dH'\over H^{\prime3}}\int_0^\infty dL\,L^5 
\cr&
\times\ \vbarhiggs^6\,(-\vhiggs)^{-(4+l)}\,\Lambda^8\,\pi^{10}\,2^6
\cr&
\times\ \left[{\ \Big(1+{\cos^2\theta\over H'}\Big)^2\ +\ {1-|\Omega'|^2\over
4H^{\prime2}}\,\sin^2 2\theta\ }\right]
\cr&
\times\ 
\epsilon_{\alpha\beta}\,S^{\dalpha\alpha}(x_1,x_0)S^{\dbeta\beta}(x_2,x_0)\,
\epsilon_{\gamma\delta}\,S^{\dgamma\gamma}(x_3,x_0)S^{\ddelta\delta}(x_4,x_0)
\cr&
\times\ 
\,G(x_5,x_0)\times\cdots\times G(x_{l+4},x_0)
\cr&
\times\ 
\Big(-{\partial\over\partial\lambda}\Big)^{l+4}\Big|_{\lambda=1}\
\exp\Big(\,  -4\pi^2\lambda 
L|\vhiggs|^2\big(\,1\,-\,|\Omega'|^2\cos^2\theta/H'\,\big)\, \Big)
\cr=\ \
\int d^4x_0&\,\epsilon_{\alpha\beta}\,S^{\dalpha\alpha}
(x_3,x_0)S^{\dbeta\beta}(x_2,x_0)\,
\epsilon_{\gamma\delta}\,S^{\dgamma\gamma}(x_3,x_0)S^{\ddelta\delta}(x_2,x_0)
\cr&
\times\ 
\,G(x_5,x_0)\times\cdots\times G(x_{l+4},x_0)\, (9+l)!\ 
{\Lambda^8\,\bigI \over\pi^2 2^6 (-\vhiggs)^{10+l}}
}}
In obtaining this equality we have performed first the trivial integration
over $L$, then the $\lambda$ differentiation.

\bf5\rm. The quantity $\bigI$ introduced in Eq.~\Gfbecomes\ stands for
the remaining scaleless 3$\hbox{-}$dimensional  integral
\eqn\bigIdef{\eqalign{\bigI\ =\ 
\int_0^1& d|\Omega'|\,|\Omega'|^6
\int_{-1}^1d(\cos \theta)\,\cos^2\theta\int_{1+2|\Omega'|}^\infty
{dH'\over H^{\prime3}}
\cr&\times\ 
{ \Big(1+{\cos^2\theta\over H'}\Big)^2\ +\ {1-|\Omega'|^2\over
4H^{\prime2}}\,\sin^2 2\theta\over
\Big(\,1\,-\,{|\Omega'|^2\cos^2\theta\over H'}\,\Big)^6}\quad .}}
It is elementary, and gives $1/48.$ 

\bf6\rm. As in the 1-instanton case we can view the 2-instanton
result \Gfbecomes\ as arising from the following effective local vertex,
built out of the total Higgs field $A(x_0)=\vhiggs+\delta A(x_0)$:
\eqn\twoIeffvertex{\sum_{l=0}^\infty\ {1\over2!\,2!\,l!}\,\psi^2\lambda^2
(\delta A)^l\
{(9+l)!\,\Lambda^8\over3\cdot2^{10}\,\pi^2(-\vhiggs)^{10+l}}\ =\
{945\Lambda^8\over32\pi^2 g^6}\,
{\psi^2\lambda^2\over A^{10}}\ .}
We have used the binomial theorem once again, and
restored the explicit $g$ dependence. Identical manipulations may
be used for the families of anomalous magnetic moment and field-strength
Green's functions, respectively \secondfamily\ and \thirdfamily, as in Sec.~5.
Comparing the fraction $945/32$
with the Seiberg-Witten effective Lagrangian, Eqs.~\FSWdef\
and \Leffcomp, we deduce for the 2-instanton coefficient
${\cal F}_2=5/16\,,$ confirming their prediction.

$$\scriptstyle{************************}$$

We are indebted to our colleagues at Durham, Los Alamos, SLAC and
Swansea for hospitality and valuable discussions while this work
was in progress. We thank T. Bhattacharya and E. Corrigan
for several enlightening conversations.
We would especially like to acknowledge A. Yung, 
for numerous valuable discussions, for access to his calculations
on the effective instanton interaction approach to these
issues \yung, and above all for spurring our interest in superinstantons.
The work of ND was supported in part by a PPARC Advanced Research
Fellowship; both ND and
 VK were supported in part by the Nuffield 
Foundation; MM was supported by the Department of Energy.
\vfil\eject
\appendix{A}{Supersymmetric and Superconformal Invariance}

In this Appendix we review some of the classical 
invariances of the microscopic
Lagrangian \Ltwodef\ together with the induced Euler-Lagrange equations 
\eulergauge{}-\eulerantimatter{}.
 To begin with, they are invariant under the following
 $N=2$ \susy\ transformations, which commute with Wess-Zumino gauge fixing:
\eqna\susytwodef
$$\eqalignno{
\delta\uv^m\ &=\ -i\ulambdabar\sigmabar^m\xione+i\xionebar\sigmabar^m\ulambda
-i\upsibar\sigmabar^m\xitwo+i\xitwobar\sigmabar^m\upsi
&\susytwodef a
\cr
\delta\ulambda\ &=\ -\xione\sigma^{mn}\uv_{mn}+i\xione\uD
+i\sqrtwo \xitwobar\Dbarslash\uA-\sqrtwo \xitwo\uF
&\susytwodef b\cr
\delta\ulambdabar\ &=\ -\xionebar\sigmabar^{mn}\uv_{mn}-i\xionebar\uD
 +i\sqrtwo\xitwo\Dslash\uAbar-\sqrtwo\xitwobar\uFbar
&\susytwodef c\cr
\delta\uD\ &=\ -\xione\Dslash\ulambdabar+\xionebar\Dbarslash\ulambda
-\xitwo\Dslash\upsibar+\xitwobar\Dbarslash\upsi
&\susytwodef d\cr
\delta\uA\ &=\ \sqrtwo\xione\upsi
- \sqrtwo\xitwo\ulambda
&\susytwodef e\cr
\delta\upsi\ &=\ -i\sqrtwo \xionebar\Dbarslash\uA+\sqrtwo \xione\uF
-\xitwo\sigma^{mn}\uv_{mn}+i\xitwo\uD
&\susytwodef f\cr
\delta\uF\ &=\ i\sqrtwo\xionebar
\Dbarslash\upsi+2ig\,[\,\uA\,,\xionebar\ulambdabar\,]\,
-i\sqrtwo\xitwobar\Dbarslash\ulambda+2ig\,[\,\uA\,,\xitwobar\upsibar\,]\,
&\susytwodef g\cr
\delta\uAbar\ &=\ \sqrtwo\xionebar\upsibar
- \sqrtwo\xitwobar\ulambdabar
&\susytwodef h\cr
\delta\upsibar\ &=\ -i\sqrtwo\xione\Dslash\uAbar+\sqrtwo\xionebar\uFbar
 -\xitwobar\sigmabar^{mn}\uv_{mn}-i\xitwobar\uD
&\susytwodef i\cr
\delta\uFbar
\ &=\ i\sqrtwo\xione\Dslash\upsibar+2ig\,[\,\uAbar,\xione\ulambda\,]\,
- i\sqrtwo\xitwo\Dslash\ulambdabar-2ig\,[\,\uAbar,\xitwo\upsi\,]\,
&\susytwodef j\cr
}$$
For purposes of comparison with Sec.~4, take $\xi_1\rightarrow \xi$ and
$\xi_2\rightarrow-\xi'.$
Actually it suffices to restrict our attention to $N=1$ invariance,
setting $\xitwo=\xitwobar=0$ henceforth, so long as we also enforce $SU(2)_\R$
invariance as per Eq.~\SUtwoR.

An heuristic route to the superconformal  group is to pose the question,
can these \susy\ transformations be made \it local\rm, that is to say
$\xi_1\rightarrow\xi(x)$ and $\xibar_1\rightarrow\xibar(x)$? The answer
involves a lengthy but straightforward calculation, in which one
repeatedly exploits not only the standard Wess and Bagger identities,
but also the following useful facts about Pauli matrices and
covariant derivatives, respectively:
\eqna\useful
$$\eqalignno{\sigmabar^m\sigma^{kl}\ &=\ \hf\big(
\eta^{ml}\sigmabar^k-\eta^{mk}\sigmabar^l-i\epsilon^{mkln}\sigmabar_n\big)
&\useful a
\cr
\sigmabar^{kl}\sigmabar^m\ &=\ \hf\big(
-\eta^{ml}\sigmabar^k+\eta^{mk}\sigmabar^l-i\epsilon^{mkln}\sigmabar_n\big)
&\useful b\cr
\sigma^m\sigmabar^{kl}\ &=\ \hf\big(
\eta^{ml}\sigma^k-\eta^{mk}\sigma^l+i\epsilon^{mkln}\sigma_n\big)
&\useful c\cr
\sigma^{kl}\sigma^m\ &=\ \hf\big(
-\eta^{ml}\sigma^k+\eta^{mk}\sigma^l+i\epsilon^{mkln}\sigma_n\big)\ ,
&\useful d}$$
and
\eqna\moreuseful
$$\eqalignno{
[\,\D_m,\D_n\,]\,{\uX}\ &=\ -ig\,[\,\uv_{mn},{\uX}\,]&\moreuseful a
\cr 
\big(\Dslash\Dbarslash\,\big)_\alpha^{\ \beta}{\uX}
\ &=\ 
-\delta^{\ \beta}_\alpha\,\D^2{\uX}\ -\ ig\,{\sigma^{mn}}_\alpha^{\ \beta}
\,[\,\uv_{mn},{\uX}\,]\,\ ,&\moreuseful b
\cr
\big(\Dbarslash\Dslash\,\big)^{\dot\alpha}_{\ \dot\beta}{\uX}
\ &=\ 
-\delta_{\ \dot\beta}^{\dot\alpha}\,\D^2{\uX}\ -\ 
ig\,\sigmabar^{mn\,\dot\alpha}{}_{\dot\beta}
\,[\,\uv_{mn},{\uX}\,]&\moreuseful c
\cr
\epsilon^{klmn}\D_k\uv_{lm}\ &=\ 0\ .&\moreuseful d}$$
In addition one needs the 
transformation law for $\uv_{mn}$ which follows from that for $\uv_m$:
\eqn\vmntrans{\eqalign{
\delta\uv_{mn}\ &=\ i\xi(x)\big(\sigma_n\D_m-\sigma_m\D_n\big)\ulambdabar
+i\xibar(x)\big(\sigmabar_n\D_m-\sigmabar_m\D_n\big)\ulambda
\cr
&
-i\ulambdabar\big(\sigmabar_n\partial_m-\sigmabar_m\partial_n\big)\xi(x)
-i\ulambda\big(\sigma_n\partial_m-\sigma_m\partial_n\big)\xibar(x)\ .
}}

The results of this exercise are as follows. Focusing first on
$\L_{\rm gauge},$ one finds invariance if and only if
\eqn\xiofxdef{\xi(x)\ =\ \xi-\etabar\sigmabar^kx_k\ ,\quad
\xibar(x)\ =\ \xibar+\eta\sigma^kx_k\ .}
The new Grassmann parameters
 $\eta^\alpha$ and $\etabar_{\dot\alpha}$ are associated with 
the superconformal generators $S^\alpha$ and $\bar S_{\dot\alpha}$, 
which are the fermionic superpartners of the special conformal generators
$K_n$ \Fayet. 
As for $\L_{\rm chiral}$, the substitution
\xiofxdef\ does not \it quite \rm work; one has to add
 an extra piece to the transformations 
for $\upsi$ and $\upsibar,$ proportional to $\eta\uA$ and $\etabar\uAbar,$
respectively.  These extra pieces are just a particular resolution of
the operator ordering ambiguity between $\D_m$ and $\xi(x)$; they are
present even in the simplest superconformally invariant model,
that of a single massless chiral superfield \Clark.
In sum, $\Ltwo$ is invariant under the space-dependent WZ-gauge-preserving
\susic\ transformations \xiofxdef,  as follows:
\eqna\sconfdef
$$\eqalignno{
\delta\uv^m\ &=\ -i\ulambdabar\sigmabar^m\xi(x)+i\xibar(x)\sigmabar^m\ulambda
&\sconfdef a
\cr
\delta\ulambda\ &=\ -\xi(x)\sigma^{mn}\uv_{mn}+i\xi(x)\uD
&\sconfdef b\cr
\delta\ulambdabar\ &=\ -\xibar(x)\sigmabar^{mn}\uv_{mn}-i\xibar(x)\uD
&\sconfdef c\cr
\delta\uD\ &=\ -\xi(x)\Dslash\ulambdabar+\xibar(x)\Dbarslash\ulambda
&\sconfdef d\cr
\delta\uA\ &=\ \sqrtwo\xi(x)\upsi
&\sconfdef e\cr
\delta\upsi\ &=\ -i\sqrtwo \xibar(x)\Dbarslash\uA+\sqrtwo \xi(x)\uF
+2\sqrtwo i\eta\uA
&\sconfdef f\cr
\delta\uF\ 
&=\ i\sqrtwo\xibar(x)\Dbarslash\upsi+2ig\,[\,\uA\,,\xibar(x)\ulambdabar\,]\,
&\sconfdef g\cr
\delta\uAbar\ &=\ \sqrtwo\xibar(x)\upsibar
&\sconfdef h\cr
\delta\upsibar\ &=\ -i\sqrtwo\xi(x)\Dslash\uAbar+\sqrtwo\xibar(x)\uFbar
-2\sqrtwo i\etabar\uAbar
&\sconfdef i\cr
\delta\uFbar\ 
&=\ i\sqrtwo
\xi(x)\Dslash\upsibar+2ig\,[\,\uAbar,\xi(x)\ulambda\,]\,&\sconfdef j}$$

Of course, given these  \it infinitesimal
\rm transformations, one  automatically knows the \it finite \rm
transformations
 as well since  exponential series in Grassmannians terminate. For instance
$\exp(\xibar\Qbar)=1+\xibar\Qbar+\hf(\xibar\Qbar)^2$ so that
\eqn\example{\exp(\xibar\Qbar)\times\uAbar\ =\ \uAbar+\sqrtwo\xibar\upsibar
+\hf\cdot\sqrtwo\xibar\big(\sqrtwo\xibar\uFbar\big)\ ;}
for our purposes
the fields on the right are then replaced by the initial ``reference''
choice of configuration. In practice, for
 superinstantons, the quadratic terms often vanish by virtue of the equations
of motion.

Finally, it is easily  checked using Eqs.~\useful{}-\moreuseful{}
that  these transformations automatically generate from the instanton
the two different types of Weyl zero modes of $\Dbarslash$: the
\susic\ modes $\xi\sigma^{mn}\uv^\cl_{mn}$, and the superconformal modes
which appear both as 
$x_k\etabar\sigmabar^k\sigma^{mn}\uv^\cl_{mn}$
and as $\xibar\Dbarslash\uA_\cl$, the relationship between them being
given by Eqs.~\etabarprimedef{}-\sconfrewrite.

\appendix{B}{Adjoint Fermion Zero-Mode Jacobian in the 2-Instanton Sector}
\def\Crbar{\bar C_r}
\def\Cs{C_s}
\def\Pinfty{{\cal P}_{\sst\infty}}
\def\tm{\tilde m}
\def\Cbar{\bar C}

Here we derive the expression \Jfermipropto\ for
 the adjoint fermion zero-mode Jacobian $J_\fermi$. Rather than work directly
with the adjoint spinor
 zero modes \lambdazm, it is equivalent but somewhat more
convenient to calculate, instead, the overlap matrix for the 
closely related adjoint \it vector \rm zero modes $Z_n$. These are
given by \Oone\
\eqn\Zmudef{Z_n\ =\ \Ubar C\sigmabar_n f\,\bbar U-\Ubar bf\sigma_n
\Cbar U}
where $C$ is an $(n+1)\times n$-dimensional matrix of constant quaternions.
The defining equations for  background-gauge vector zero modes, namely
\eqn\veczmdef{\D_{[\,n}\,Z_{m\,]}
\ =\ \big(\D_{[\,n}\,Z_{m\,]}\big)^{\rm\sst dual}\ ,\qquad
\D_n Z^n\,=\,0\ ,}
ensure that, when they are viewed as infinitesimal transformations
on the gauge field, the field strength remains self-dual. The
first of these conditions leads to 
$\Deltabar C-(\Deltabar C)^T=-\Cbar\Delta+(\Cbar\Delta)^T$
while the second (the background gauge condition) works out to
$\Deltabar C-(\Deltabar C)^T=\Cbar\Delta-(\Cbar\Delta)^T$; combining
these gives 
$\Deltabar C=(\Deltabar C)^T$, or equivalently
\eqn\linconst{\abar C\,=\,(\abar C)^T\ \ ,\qquad\bbar C\,=\,(\bbar C)^T\ .}
These are the precise analogs of the linear constraints
on the fermion zero modes, Eqs.~\zmconsone-\zmconstwo; indeed the fermion
zero modes can be linearly obtained from the vector zero modes by folding in
a spin matrix times a spinor.

That a calculation of the overlap matrix between different $Z_{rn}$ 
 is at all tractable is thanks to a remarkable identity due to
Corrigan:\foot{The
index $r$ labels the different zero modes. This
identity is quoted in Osborn as his Eq.~(3.17), but its proof is apparently
not to be found in the literature; we supply one in Appendix~C.}
\eqn\corrid{\trtwo Z_{rn}Z_s^n\ =\ \hf\,\partial_n\partial^n
\Tr\,\Crbar(\P+1)\Cs f\ ,}
where the capitalized
`Tr' means a trace over both $SU(2)$ and ADHM indices. The inner product
is then,
\eqn\appcone{\eqalign{\big\langle Z_r\big|Z_s\big\rangle\ &\equiv\ 
-\int d^4x\,\trtwo Z_{rn}Z_s^n\ =\ -\hf\int d^4x\,\partial_n\partial^n
\,\Tr\,\Crbar(\P+1)\Cs f
\cr&=\ -\pi^2\lim_{r\rightarrow\infty}\,r^3\,\Tr\,\Crbar(\Pinfty+1)\Cs
f'(r)
\cr&=\ 
2\pi^2\,\Tr\,\Crbar(\Pinfty+1)\Cs\ .}}
Here
\eqn\Pinftydef{\Pinfty+1\ \equiv
\ \lim_{r\rightarrow\infty}\big(\, 2-\Delta f\Deltabar\,)\ =\ 2-b\bbar
\ =\ \pmatrix{2&0&0\cr0&1&0\cr0&0&1}\ ,}
specializing at last to the 2-instanton case.
In obtaining these expressions we have used the asymptotic forms
\asymadhm{a,b}. 
Also we have disregarded the term where the normal 
derivative hits $\P$ rather than $f$, as this costs two powers of $r$. 

We parametrize the quaternion matrix $C$ as 
\eqn\bigCdef{C\ =\ \pmatrix{m_1&m_2\cr M_3&M_1\cr M_1&-M_3} \ ,  }
 omitting the
translational modes $\propto b$ since they do not mix with the others.
Thus
\eqn\Cprod{\Tr\,\bar C\cdot(\Pinfty+1)\cdot C\ \propto\ 
|m_1|^2+|m_2|^2+|M_3|^2+|M_1|^2\ .}
Thanks to the constraints, $M_1$ can be eliminated, just like
its fermionic counterpart $\M_1$ in Eq.~\Monedef. It is convenient, first,
to define the rotated quaternionic variables $\tm_1$ and $\tm_2$, via
\eqn\tmdef{m_1\ =\ {w_2\abar_1\tm_1\over\sqrt{|w_2|^2|a_1|^2}}
\ \ ,\qquad
m_2\ =\ {w_1\abar_1\tm_2\over\sqrt{|w_1|^2|a_1|^2}} \ .}
Since  $|m_1|^2+|m_2|^2=|\tm_1|^2+|\tm_2|^2$ this change of variables
does not affect the determinant. The constraint on $M_1$ is then
resolved as:
\eqn\emonedef{\eqalign{M_1\ &=\ {a_3\over 2|a_3|^2}\,\big(2\abar_1M_3
+\wbar_2m_1-\wbar_1m_2\big)\cr&=\ 
{a_3\abar_1\over2|a_3|^2\sqrt{|a_1|^2}}\cdot
\big(\,2\sqrt{|a_1|^2}\,M_3+\sqrt{|w_2|^2}\,\tm_1
-\sqrt{|w_1|^2}\,\tm_2\,\big)\ .
}}
{}From Eqs.~\Cprod\ and \emonedef, it is obvious that the matrix
whose determinant we want can be written as
\eqn\detdef{\,\pmatrix{1&0&0\cr0&1&0\cr0&0&1\cr}\  +\ Q\cdot Q^T\, \ ,}
where
\eqn\QTdef{Q^T\ =\ {1\over2\sqrt{|a_3|^2}}\big(\
2\sqrt{|a_1|^2}\ ,\ \sqrt{|w_2|^2}\ ,\ -\sqrt{|w_1|^2}\ \big)\ .}
This matrix has two eigenvalues equaling unity (spanning the $\perp$-space
of $Q$), and the third eigenvalue, hence the determinant itself,
equaling 
\eqn\Qsqdef{1+|Q|^2\ = \ {H\over4|a_3|^2}\ ,}
with $H$ as in Eq.~\paramsdef{}. Actually this is the determinant
for \it each \rm of the four decoupled (thanks to \tmdef)
$\sigma_n$ components of the quaternion, so that
\eqn\Jfermifinal{J_\fermi \ \propto\ {H^4\over|a_3|^8}\ ,}
confirming Eq.~\Jfermipropto.

\appendix{C}{Vector Zero Mode Identities}

\subsec{An alternative route to the classical adjoint Higgs}

One important example of a vector zero mode (cf.~Eqs.~\Zmudef, \veczmdef,
and \moreuseful{a}\ ) is $\D_nA_\cl$.
 Here we confirm that our solution for $A_\cl$ does
indeed have this property. In performing this check, we are giving in effect
an alternative route to the construction of $A_\cl.$ 

We focus on the right-hand side of Eq.~\covderiv, and once again posit that
 $\A$ is a constant $(n+1)\times(n+1)$-dimensional matrix of the form shown
in Eq.~\blockdiag, in particular that $\A'$  satisfies the condition
\Asinglet. We can then reexpress
\eqn\appdone{\eqalign{\A\Delta\ &=\ 
\pmatrix{\vhiggsa w_1&\cdots&\vhiggsa w_n
   \cr{}&{}&{}\cr
 {}&\A'a'&{} \cr{}&{}&{}}
\ +\
\pmatrix{0 &{}&\Uparrow &{}
 \cr \vdots & {} & bx &{}\cr 0&{}&\Downarrow&{}}
\pmatrix{0&\cdots&0\cr{}&{}&{}\cr
 {}&\A'&{}\cr {}&{}&{}}
\cr&\equiv\
\pmatrix{\vhiggsa w_1-w_k\A'_{k1}&\cdots&\vhiggsa w_n-w_k\A'_{kn}
\cr
{}&{}&{}\cr
{}&\big[\,\A'\,,\,a'\,]&{}\cr
{}&{}&{}  }\ ,}}
with $a'$ as in Eq.~\bcanonical.
In the final rewrite we have exploited the fact
 that the matrix product is to be
left-multiplied by $\Ubar_\lambda,$ and have therefore used Eq.~\perpspace\
to eliminate the $x$ dependence. We can now check that this final
matrix satisfies the two conditions \linconst\ for a vector zero mode
matrix $C$. The latter is automatically satisfied when $\A'$ obeys
Eq.~\Asinglet\ (and is therefore necessarily antisymmetric in
ADHM indices since $\bar\A'=-\A'$). 
Less obviously, the former  is equivalent to Eq.~\defineAcl; verifying
this claim is entirely straightforward, once one rewrites Eq.~\crucial{a}
in terms of the submatrix $a'$. 

By dotting the vector zero mode $\D_nA_\cl$ with $\Thetabar\sigmabar^n$
we of course construct the superconformal fermionic zero modes 
\sconfrewrite, which are among the lifted fermionic modes in the problem.
Alternatively these modes may be built up from the field strength;
cf.~Sec.~4.2 and Appendix A.

\subsec{Proof of Corrigan's inner product formula}
\def\Csbar{\bar C_s}
\def\Cr{C_r}
Next we supply a proof of Corrigan's remarkable identity, Eq.~\corrid.
First we expand  the left-hand side, using Eqs.~\Zmudef, \projdef,
and \newfact:
\eqn\corrlhs{\eqalign{\trtwo Z_{r{m}}Z_s^{m}\ &=\ 
2\,\Tr\,\Big(\ \big(\Csbar\P\Cr\,+\,\Crbar\P\Cs\big)f\,\bbar\cdot
\trtwo\P\cdot bf
\cr&\qquad+\ \Crbar\P bf(\P\Cs)^Tbf\,+\,\P\Cr f\,\bbar(\Csbar\P)^Tf\,\bbar
\ \Big)\ .}}
In order to differentiate the right-hand side of Eq.~\corrid\ twice,
one needs, in addition to Eq.~\fid, the following useful facts:
\eqna\morefid
$$\eqalignno{\partial_{m}\partial^{m} f\ &=\
4f\,\bbar\cdot\trtwo\P\cdot bf &\morefid a
\cr
\partial^{m}\P\ &=\ -\Delta f\,\bbar\sigmabar^{m}\P-\P b\sigma^{m} f\Deltabar
&\morefid b
\cr
\partial_{m}\partial^{m} \P\ &=\ 4\{\,\P\,,\,bf\,\bbar\,\}
-4\Delta f\,\bbar\cdot\trtwo\P\cdot bf\Deltabar\ . &\morefid c}
$$
So the right-hand side of Eq.~\corrid\ becomes:
\eqn\corrhs{\eqalign{\Tr\,\Big(\ 2&\Crbar\,\{\,\P\,,\,bf\,\bbar\,\}\Cs f
+2\Crbar(\P+1)\Cs f\,\bbar\cdot\trtwo\P\cdot bf
\cr
-2&\Crbar\Delta f\,\bbar\cdot\trtwo\P\cdot bf\Deltabar\Cs f
+\Crbar\Delta f\,\bbar\sigmabar^{m}\P\Cs f\partial_{m}(\Deltabar\Delta)\,f
\cr
+&\Crbar\P b\sigma^{m} f\Deltabar\Cs f\partial_{m}(\Deltabar\Delta)\,f
\, \Big)\ .}}
Comparing Eqs.~\corrlhs\ and \corrhs, it is clear that we need to
eliminate all explicit factors of $\Delta$ and $\Deltabar$ in favor
of $\P$, using Eq.~\Deltadef. In fact the third, fourth and fifth terms
in Eq.~\corrhs\ can individually be rewritten in this manner, yielding,
respectively,
\eqn\corrhstwo{2\,\Tr\,\Big(\
\Crbar(\P-1)\Cs f\,\bbar\cdot\trtwo\P\cdot bf+
\Cr f\,\bbar(\Csbar\P)^Tf\,\bbar(\P-1)
+(\P\Cr)^Tbf\Csbar(\P-1)bf\ \Big)}
In proving this last step it is helpful to use $\Cbar\Delta=
-\Deltabar C+\trtwo\Cbar\Delta$ as follows from Eqs.~\linconst.
Now the ``$-1$'' pieces in the last two terms in \corrhstwo\ cancel the
anticommutator piece in \corrhs; by inspection,
the surviving terms establish the claimed equality with \corrlhs, \sl QED\rm.

\appendix{D}{Residual Discrete  Symmetries in the ADHM Measure}

The purpose of this Appendix is to determine the relevant 
discrete symmetry 
groups $G_{n}$ and the corresponding symmetry factors ${\cal S}_{n}$
for the case $n=1$ and the particular parametrization of the $n=2$
solution described in the text. 

After fixing the canonical form \bcanonical, the collective coordinates of the
ADHM multi-instanton live in the vector $\vW=(w_1,\cdots,w_n)$ and the
$n \times n$ submatrix ${a'}$. To solve the constraint we need to
specify $\vW$ and ${a'}$ in terms of a set of
unconstrained parameters $\{ X_{i}\}$, $i=1,2,\ldots, 8n$. 
The residual  redundancy is then obtained by finding all matrices
$R\in O(n)$ satisfying the condition \condition\ discussed in the text.
In general the solutions of this condition will form a discrete
subgroup of $O(n)$, the symmetry group $G_{n}$. The number of such
solutions is the symmetry factor ${\cal S}_{n}$. Notice that the
symmetry group and symmetry factor depend not only on $n$ but also on the
details of the particular parametrization $\{X_i\}$ chosen.

{\bf n=1:} In the ADHM construction, the single instanton solution is
parametrized as in Eq.~\oneI; in this case 
the residual  transformations are simply generated by $R=\pm 1$.
Hence the symmetry group is just $Z_{2}$ and ${\cal S}_{1}=2$. 

{\bf n=2:} Here we have 
\eqn\nicksnewmat{\vW\ =\ (w_{1},w_{2})\ ,\qquad
{a'}\ =\ \pmatrix{  x_{0}+a_{3} & a_{1}  \cr
a_{1}  & x_{0}-a_{3} }
\ .}
The constraint is solved by eliminating the parameter $a_{1}$ 
via the relation \aonedef. We
complete the parameterization by specifying $\Sigma\equiv 0$ so that
\eqn\gauge{
a_{1}={1\over4|a_{3}|^{2}}a_{3}(\bar{w}_{2}w_{1}-\bar{w}_{1}w_{2}) \ .}

Now let us solve the condition \condition. This is 
equivalent to finding the matrices $R$ such that the transformed collective
coordinates  still obey the condition \gauge.
By definition any matrix $R\in O(2)$ must have ${\rm det}R=\pm 1$. 
If ${\rm det}R=+1$, $R$ is just a rotation matrix,
\eqn\rotting{R=R_{\theta}=\pmatrix{  \cos\theta & \sin\theta \cr
-\sin\theta & \cos\theta }\ ,\qquad0\le\theta<2\pi\ ,}
and the transformed parameters from \trans\ may be rewritten as
\eqn\wtransforms{
(w^{\theta}_{1},w^{\theta}_{2})=(w_1,w_2)\cdot R_\theta\ ,\qquad
(a^{\theta}_{3},a^{\theta}_{1})=(a_3,a_1)\cdot R_{2\theta}\ ,\qquad
x_0^\theta=x_0^{}\ .}
Hence the equation we have to solve is
\eqn\gaugetheta{
a^{\theta}_{1}\ =\
{1\over4|a^{\theta}_{3}|^{2}}\,a^{\theta}_{3}
(\bar{w}^{\theta}_{2}w^{\theta}_{1}-\bar{w}^{\theta}_{1}w^{\theta}_{2})\ .}
Since
$\bar{w}^{\theta}_{2}w^{\theta}_{1}-\bar{w}^{\theta}_{1}w^{\theta}_{2}=
\bar{w}_{2}w_{1}-\bar{w}_{1}w_{2}$, this simplifies to
\eqn\newappdee{0\ =\ \abar_3^\theta a_1^\theta-\abar_3a_1\ =\
\hf\big(|a_3|^2-|a_1|^2\big)\sin4\theta\,-\,
(\abar_3a_1+\abar_1a_3)\sin^22\theta\ .}
Noting that the last term vanishes by virtue of Eq.~\gauge, and
excepting the special case $|a_{1}|=|a_{3}|$ 
discussed below, we see that the solutions with $\det R=+1$ are given by 
$\sin4\theta=0,$ i.e., $\theta=\theta_{k}=(k-1)\pi/4$ with $k=1,2,\ldots, 8$.
 Our analysis is completed by considering 
the $O(2)$ matrices with ${\rm det}R=-1$. These 
 can be written as the product of a rotation, $R_{\theta}$, and a
reflection, $R_{r}=\sigma_{3}$. The analysis proceeds as
before and yields the same values of $\theta$. 
Hence we have $16$ solutions of Eq.~\condition\ in all:
\eqn\sixteensolns{
R=R_{\theta_k}  \qquad{} {\rm and} \qquad{}  R=R_{r}R_{\theta_{k}} \ .}
It follows that ${\cal S}_{2}=16$. Because $R_{r}$ does not commute with 
the $R_{\theta_{k}}$,  $G_{2}$ is the dihedral group $D_{8}$,
 the symmetry group of a regular octagon under reflections and
rotations. 

Some particular symmetries of the two instanton solution are:
\hfil\break
\bf1\rm.
The transformation generated by $R_{r}R_{3\pi/2}$ permutes the
labels $1$ and $2$. In the clustering limit this has the
effect of interchanging the two well-separated instantons.
\hfil\break
\bf2\rm.
 The matrix $R_{r}R_{3\pi/4}$ 
has the effect of interchanging the parameters $a_{1}$ and
$a_{3}$. As described in the text, the parameter $a_{3}$ has
the interpretation of (half) the distance between two well separated
instantons in the clustering limit: $|a_{3}|\rightarrow \infty$. The 
discrete 
symmetry means that the region of the parameter space $|a_{3}|<|a_{1}|$ 
is equivalent to the region $|a_{3}|>|a_{1}|$. Hence 
the limit $|a_{3}|\rightarrow 0$ is also a  clustering limit. The fixed
points of the discrete symmetry at $|a_{1}|=|a_{3}|$ are the central
points in the moduli space at which the two instantons are coincident.
Reasoning by analogy with the Jackiw-Nohl-Rebbi parametrization
discussed by Osborn (see Sec.~4 of Ref.~\Oone), we expect that
at these central points
the solution degenerates to one of lower topological charge $n=1$.

\listrefs
\bye